# Economic Reality, Economic Media and Individuals' Expectations

Kristoffer Persson

15 January 2020

**Abstract**

This paper investigates the relationship between economic media sentiment and individuals' expectations and perceptions about economic conditions. We test if economic media sentiment Granger-causes individuals' expectations and opinions concerning economic conditions, controlling for macroeconomic variables. We develop a measure of economic media sentiment using a supervised machine learning method on a data set of Swedish economic media during the period 1993–2017. We classify the sentiment of 179,846 media items, stemming from 1,071 unique media outlets, and use the number of news items with positive and negative sentiment to construct a time series index of economic media sentiment. Our results show that this index Granger-causes individuals' perception of macroeconomic conditions. This indicates that the way the economic media selects and frames macroeconomic news matters for individuals' aggregate perception of macroeconomic reality.

Kristoffer Persson, Lund University School of Economics and Management, Box 7082, 220 07 LUND, Sweden

*Correspondence: kristoffer.persson@nek.lu.se*

Acknowledgements: Funding from the Jan Wallander and Tom Hedelius Foundation is gratefully acknowledged.

# 1. Introduction

Individuals gather information on economic conditions from several different sources. Some of the main sources are various media outlets, such as newspapers, magazines and online publications (Fogarty 2005). The media often reports on current factual events and, to some extent, re-reports news that is published by experts in public reports and scientific papers (Nadeu *et al.* 1999). However, the media is not entirely neutral and can sometimes put a filter on reality when newspapers select and frame news stories to capture individuals' attention (Gentzkow & Shapiro 2008; McCarthy & Dolfsma 2014). Individuals using the media to inform themselves about economic conditions may thus become biased in their opinions and expectations about the future, causing them to make wrong decisions (van Raaij 1989).

Public opinion about economic conditions is paramount to economic outcomes. Individuals' expectations about the state of the economy influence their decisions in terms of savings, consumption, investments and entrepreneurial endeavours (*e.g.* Keynes 1936; Lachmann 1943; Muth 1961, Arenius & Minniti 2005; Koellinger *et al.* 2007; Brown & Taylor 2006; Moretti 2011; Hirshleifer 2001; Baker & Wurgler 2007; Burnside *et al.* 2011). Because the media has an incentive to put a filter on economic conditions, they may have a de-stabilizing effect on the economy. For example, since the media tries to capture individuals' attention by over-emphasizing the negative aspects of economic conditions, they may cause individuals to overestimate the risk of the economy entering into a recession. As a consequence of this individuals may decide to postpone consumption, investment and entrepreneurial endeavours more than what is motivated by actual economic conditions. This could potentially be the decisive factor that causes the economy to go into a recession (Pigou 1927; Jaimovich & Rebelo 2009).

In this paper, we test if the way Swedish media report on economic conditions affects individuals' opinions and expectations about them. We do this by testing if individuals' opinions and expectations about economic conditions are Granger-caused by economic media sentiment when controlling for macroeconomic variables. We develop a measure of media sentiment using a supervised machine learning method for text classification of media items' sentiment based on their full textual content. We use 179,846 media items, stemming from 1,071 unique media outlets including the major nationwide newspapers, local newspapers and online news during the 1993–2017 period, to observe the development of sentiment in the Swedish economic media. We are thus including several booms and busts in the real economy and in the financial markets. The data that we use gives us unique possibilities compared to other studies that typically are limited to a few time periods or a few media outlets.

Our results show that the economic media sentiment is slightly positive on average. Individuals' opinion about current macroeconomic conditions is Granger-caused by economic media sentiment and affected by macroeconomic variables. However, economic media sentiment does not Granger-cause individuals' expectations about future macroeconomic conditions, individuals' expectations about their own personal economic situation or individuals' evaluation of current developments concerning their own private economic situation.

The rest of the paper is outlined as follows. Section 2 develops some hypotheses concerning the effects of competition among media companies on economic media sentiment and the effect of economic media sentiment on individuals' opinions and expectations. Section 3 presents economic media data, section 4 presents macroeconomic- and expectations data. Section 5 presents results and section 6 concludes the paper.

## 2. The News Media and Competition: Three Hypotheses

Information published by the media is an important component of an individuals' information set. Individuals are constrained in their attention and thus often use the media to form an opinion about matters that require high cognitive effort, such as predicting economic outcomes. Information presented by the media is easier for individuals to access and process than, for example, economic and meteorological data published in tables or reports by national statistics agencies and central banks. Individuals who consume information published in the media thus only need to use relatively little cognitive effort to form an opinion about complex matters, which allows them to allocate their attention efficiently. The way the media reports economic conditions may thus affect individuals' information set and thereby the decisions they make (*e.g.* Carroll 2003, Sims 2001).

In most economies, different media outlets compete in a market for news (Winseck 2008). Competition for market shares between different media outlets should theoretically cause the media to report on events that are of great importance to most individuals (Mullainathan & Schleifer 2005). Media that are successful in doing so in a fast, accurate and neutral manner will also be successful in selling their product and in generating returns to their shareholders. If the media reports relevant information, this will enable individuals to be well informed about current and important matters, which will allow them to make adequate decisions (Gentzkow & Shapiro 2008).

Competition for market shares between different media outlets may, however, cause the media to not report objectively on economic conditions. The reason for this is that when media companies compete for market shares they are *de facto* competing for consumers' *attention*. This may cause the media to exaggerate the importance of some events and underemphasize the importance of others. The daily press, for example, may potentially maximize its revenue by selecting their main stories based on how well the essence of the story can be translated into an eye-catching headline. Such a headline has a greater potential of capturing individuals' attention, causing them to buy single copies of the newspaper (Tannenbaum 1953). The media may thus not report on economic conditions to the extent that is actually motivated by economic data, but instead report more on other topics which the media regard as better headline material (Andrew 2007). Another possible consequence of media competition is that the media may over-report and exaggerate the consequences of negative aspects of economic conditions. This implies that the media, albeit reporting on relevant and important economic events, tend to exaggerate the sensational and eye-catching aspects of economic events such as economic crisis and stock market crashes in order to attract individuals' attention (Hester & Gibson 2003). Newspapers may also capture the attention of customers by tailoring their reporting to fit the political-ideological conviction of their target market segment. Newspapers thus over-emphasize some topics, such as unemployment, if the opposing political party is in power (Larcinese *et al.* 2011).

There is some evidence in the literature showing that the media sentiment is not neutral. The media generally reports on current events and we thus expect economic media to mainly report on current business cycle conditions rather than long-term economic developments, such as economic growth. A neutral economic media should thus publish equally many media items with positive and negative sentiment, in the long run. Most previous studies, however, find that the media tends to publish more content with negative sentiment than positive sentiment, which may be a consequence of the competition between media outlets (Fogarty 2005; Arango-Kure *et al.* 2014; Boydstun *et al.* 2018). This literature, however, mainly uses US data and few studies have investigated whether media in other economies behave in a similar way. Media non-neutrality is likely to exist also in Sweden where the media, for example, reported heavily on the negative consequences of the global financial crisis of 2008, despite the fact that economic data showed that the Swedish economy was less affected by the crisis than many other economies (Claessens *et al.* 2010; Olsson *et al.* 2015). This brings us to our first hypothesis:

**H1: Economic media is not neutral and therefore publishes unequally many items with positive and negative sentiment.**

Since individuals use the media to inform themselves about economic conditions, they may become biased in their opinions and expectations about economic conditions (van Raaij 1989). There is, however, no consensus in the literature concerning the effect of economic media on individuals' evaluations and expectations about economic conditions. Some studies find little or no effect of the media on individuals' opinions about current and future economic conditions (Fogarty 2005). Others find that the economic media affects individuals' opinions about both the current state of the macro economy and their expectations about the future developments of macroeconomic conditions (Hester & Gibson 2003). There is also some evidence that media sentiment that cannot be explained by economic data affects individuals' opinions and expectations about economic conditions. This suggests that media non-neutrality may influence individuals' perceptions of economic reality (Boydstun *et al.* 2018). This brings us to our second hypothesis:

**H2: Individuals' opinions and expectations about current and future economic conditions are Granger-caused by economic media content that is not explained by economic data.**

Individuals are sometimes better informed than the media about changes in economic conditions. The economic media typically reports on changes in macroeconomic conditions that affect most individuals in the economy. Individuals may, however, have access to better sources of information about their own personal economic situation that the media cannot access (Wåhlberg & Sjöberg 2000). This information is likely to be private and more directly linked to individuals' own economic situation thus causing them to use it instead of economic media information. Opinions about their own personal economic situation for individuals that have good unemployment insurance are, for example, not likely to be affected if the media reports that a recession is coming, compared to individuals without unemployment insurance.

Individuals may also disregard media information about the macro economy because they suffer from the illusion of control bias. They are thus overconfident about their own economic situation in relation to the macro economy and may therefore underestimate individual level economic risk, such as the risk of becoming unemployed. Individuals may thus disregard media reporting about negative macroeconomic events when they assess their own personal economic situation (Bovi 2011). This brings us to our third and final hypothesis:

**H3: Economic media content does not Granger-cause individuals' opinions and expectations about their private economic situation.**

## 3. Economic Media Data

In order to test our hypotheses, we need to quantify economic media content. Some studies do this by using economic data as proxies for economic media content (Alwathainani 2010; Nguyen & Claus 2013). This approach is problematic since it is not a direct measure of media content. It thus assumes both that individuals use the media to inform themselves about economic conditions and that the media reports neutrally on these conditions. These assumptions are highly restrictive as they do not allow individuals to be affected by both the media and by economic data, and because they do not allow for non-neutral media content. Other studies quantify media content by classifying media items into classes based on sentiment, *i.e.*, whether the texts have a positive or negative tone. One approach is to classify the headlines and economic content on the front pages of influential newspapers with respect to their sentiment (Hester & Gibson 2003; Doms & Morin 2004; Birz & Lott 2011). This approach relies heavily on the assumption that media content can be reduced to headlines and front-page material without any substantial loss of sentiment information. Such an assumption is not likely to hold in general, because headlines and front pages are tools that media outlets use to capture individuals' attention and thus differ from media items' actual sentiment (Andrew 2007).

We follow a strand in the literature that classifies media items' actual textual content (Nadaeu *et al*. 1999; Fogarty 2005; Soroka 2006; Goidel *et al*. 2010; Lamla & Maag 2012; Lamla & Lein 2014; Dräger 2015; Baker *et al*. 2016; Armelius *et al*. 2017; Boydstun *et al*. 2018; Lamla *et al*. 2019). We classify economic media items' full textual content into positive and negative sentiment classes, which we then use to quantify media content by counting the number of media items in each month that belong to the positive and negative classes. We are thus able to create a time series index of economic media sentiment. The process by which we create this index is comprised of six individual steps as shown by Figure 1. In the first step, we download economic media items from an online media database. In step 2, we parse the downloaded information by separating media items' content from each other and by separating media items' header, body, publisher and date of publication into separate pieces of information; in step 3 we create test data by manually classifying a sample of news items with respect to sentiment. In step 4 we estimate a probability model using the test data we created in step 3. In step 5 we classify all of the media items that we have downloaded using the probability model from step 4 and in step 6 we calculate our economic media sentiment index.

**Figure 1. Workflow for Quantifying Media Sentiment**

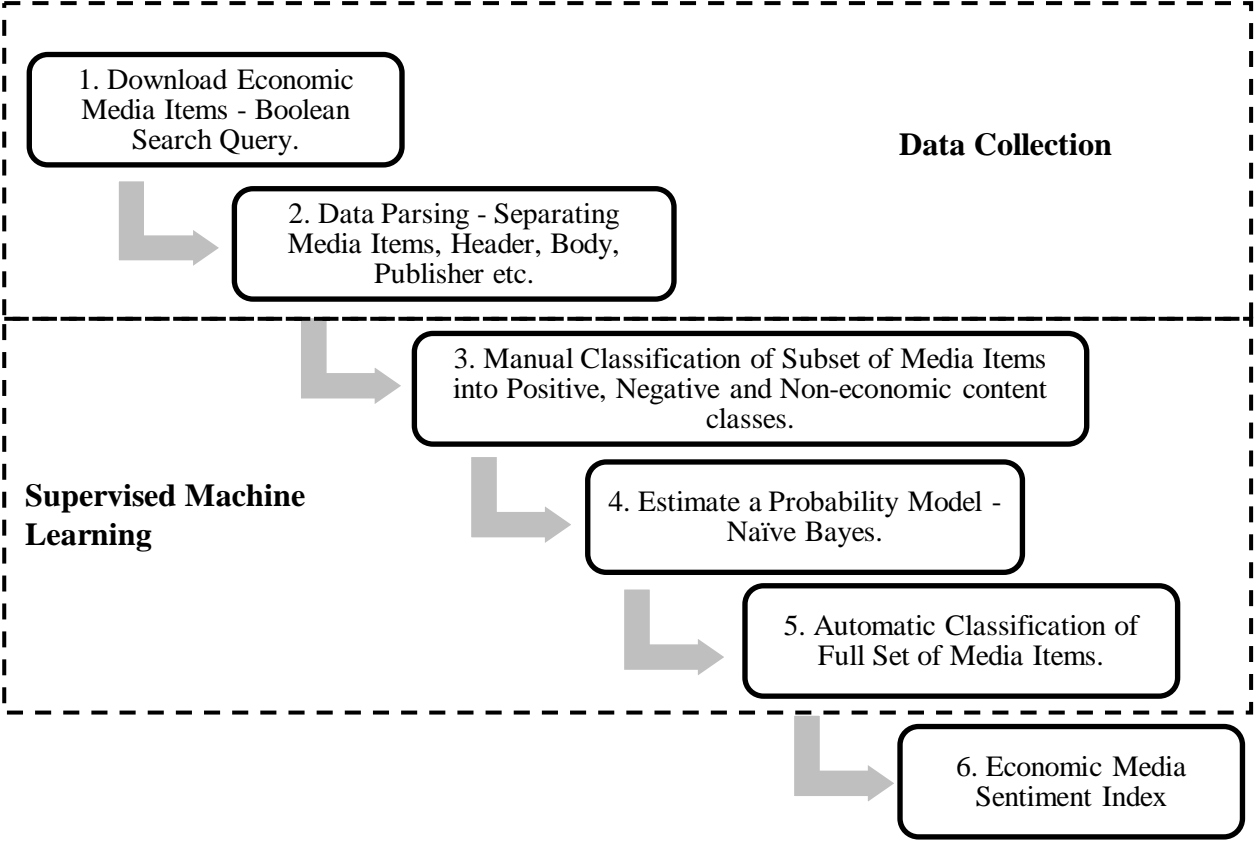

## 3.1 Data Collection

We use media data from the Retriever database, which contains most editorial media items published in print and online in Sweden since the 1980s. The database contains around one hundred million searchable media items published by a wide range of media outlets including all the major newspapers. Before the year 2000, the database consisted primarily of printed newspapers with nationwide coverage. From the year 2000 and onwards, the database also includes media items published online. This caused the number of media outlets in the database to increase and changed the composition of media sources to include more local daily newspapers, purely Internet based news sites, weekly news magazines, trade magazines, union magazines and public authorities. The data contains information on whether media items are published in print or online which allows us to split the data accordingly. We are thus able to control for the effects of the influx of online media into the database in our analysis.

We are interested in downloading as many media items with economic content as possible. We thus use a search query[1] in the Retriever search engine, which is based on an economic dictionary we have downloaded from [www.ekonomifakta.se.](www.ekonomifakta.se.) This webpage presents information and data on matters of importance to the Swedish economy and is run by the Confederation of Swedish Enterprise (*Svenskt Näringsliv*) which is Sweden's largest employer organization. The dictionary contains 678 words and concepts used in the fields of economy, taxes, the labour market, energy, environment, entrepreneurship, education and the public sector. To download as many media items with economic content as possible from the Retriever database, we created our search query by requiring that at least one of the words in the dictionary is present in the media items' texts. However, this may cause us to download a substantial number of media items that do not have economic content due to the comprehensiveness of the *Ekonomifakta* dictionary. We thus remove those items that do not have economic content at a later stage when we classify our downloaded media items using supervised machine learning.

One potential problem that arises when using the *Ekonomifakta* dictionary is that some words do not have explicit economic meaning when taken out of context. For example, if an environmental word occurs in a text, without any other economic words being present, this would not be a good indication that the media item has any economic relevance. Another potential issue with our search query is that the economic terms in the dictionary may give hits for very short economic media items that merely report on one topic without any sentiment. Finally, the query may result in media items that report economic matters in a naïve way that lacks realism compared to what the economic data actually suggests. Short newspaper notes may, for example, report that inflation is increasing or falling without presenting any information on other economic variables such as unemployment or GDP. Such media items do not contain any sentiment because higher and lower inflation can be both negative and positive for the economy. Chronicles published in small online publications may, on the other hand, report on a plenitude of economic conditions. These media items may report on economic conditions in a naïve way compared to other media items that are published by the larger and better funded media outlets. The sentiments of these naïve media items may thus make our measure less precise since they are likely to deviate from what the economic data actually suggests.

We mitigate these issues by augmenting our search query to require that the word "economy" ("*ekonomi*" in Swedish) is present in the texts together with either of the words "forecast" ("*prognos*" in Swedish) or "report" ("*rapport*" in Swedish). This increases the probability that the search query returns media items that report on several economic conditions in the same text, since they will be reporting on findings and statements stemming from reports and forecasts made by economic bodies such as commercial banks and the central bank (*Riksbanken*). Such reports and forecasts are typically based on the outcomes of econometric models using economic data, and are thus well anchored in economic reality. These documents report general economic conditions, albeit with a main focus on one economic variable such as inflation or the interest rate (Blinder *et al*. 2001).[2] Another advantage of this approach is that we ensure that our media items are especially suited for testing Hypotheses 2 and 3. Here, we use consumer sentiment data on individuals' retrospective evaluations of recent economic outcomes and their prospective expectations about future economic conditions. There is thus symmetry between our media data and the consumer sentiment data in the sense that media items containing the word "forecast" are prospective as they primarily deal with future economic developments, and media items containing the word "report" are retrospective in the sense that they primarily deal with realized economic outcomes.

Some economic events, such as economic crises, have greater impact on the economic media than other events. The Retriever search engine allows us to measure the impact of economic events on the economic media because many media items reporting a particular event are re-reported by several other media outlets. A typical example of this is news items reported by *Tidningarnas Telegrambyrå* (TT) that have a big impact on the news media as a whole in times of major economic events such as economic crisis. This is especially true in the initial stages of unravelling events, when many smaller daily newspapers struggle to keep up with the larger and better-funded newspapers.

Our search query resulted in 179,846 media items published in print and online. The Retriever home page delivered the data in text files in batches of max 500 media items per file. We parsed the information in these text files into separate pieces of information so that we could separately handle the media items' header, body of text, source, publication date/time and place of publishing (print/online). This allows us to split the data based on the parsed information and conduct our analysis on different subsets of the data.

### 3.2 Supervised Machine Learning

Supervised machine learning (SML) is a set of techniques often used to classify large bodies of data. Typical applications of SML are email spam filters (emails being spam or non-spam), topic-specific search engines (document topics, *e.g.*, economics, politics, sports, *etc.*) and detecting sentiment in consumer reviews (Manning *et al.* 2010). SML techniques are often deployed in situations where manual classification of data is not possible due to time and resource constraints. These techniques have received limited attention by researchers in the field of economic media sentiment detection, with some promising exceptions (*e.g.* Boydstun *et al.* 2018).

Most of the SML-techniques used for text classification work on training data, which consist of text items that are pre-classified by humans, to estimate a probability model. These SML-techniques use the estimated probability model for out-of-training-data classification, *i.e.*, to assign a class to text items that were previously unseen by the probability model. One can evaluate the accuracy of the probability model by partitioning the training data into a training set, which is used to estimate the probability model, and a validation set, which is disjoint with the training data set and is used to evaluate the accuracy of the models' predictions. This process is known as K-fold cross validation (Sebastiani 2002), and can be used once or several times.

**3.2.1 Training Data**

We randomly select a set of news items from the economic media items we have downloaded, and manually classify them into three groups: (i) positive sentiment, (ii) negative sentiment or (iii) irrelevant.[3] This process results in 31 % of the media items in the test data being classified as positive sentiment, 29 % classified as negative sentiment and 40 % classified as irrelevant. The relatively high number of irrelevant media items in the test data is most likely due to the fact that we use a comprehensive search query to retrieve our data from the Retriever database. This class consists of media items that do not have any economic content. A frequent kind of irrelevant media item in the test data are TV-guides that end up in the search results because they contain an occasional word from the search query.

We classify documents as negative sentiment if, for example, they report on the consequences of economic crises, rising unemployment and/or a slowdown in GDP growth. A typical example of a media item which we classify as negative is a report of a forecast from a commercial bank stating that the economy is losing its momentum or that a trade organization for the construction industries has released a report stating that building investments are down. We classify documents as positive sentiment if, for example, they report falling unemployment, increased GDP growth or general improvements of business cycle conditions.[4] A typical example of a media item which we classify as positive economic sentiment is a report from the Swedish National Institute for Economic Research saying that investments are growing in the economy, or that a commercial bank issued a forecast which predicts a recovery in the economy. However, media items do not have to report directly about macroeconomic business cycle conditions to be classified as positive or negative sentiment. If, for example, a media item reports that a major company is laying off (hiring) workers due to a decrease (increase) in demand, then we classify this as negative (positive) economic sentiment. Such actions taken by individual firms contribute to the general business cycle conditions via an increase (decrease) in unemployment.

We take into account the fact that Sweden is a small and export-dependent economy which is sensitive to changes in foreign demand for goods and services (Gylfason 1999). We do this by classifying economic media items that report on improved (worsened) economic conditions in foreign economies as positive (negative) sentiment, since increases (decreases) in demand in foreign economies are likely to cause an increase (decrease) in the export of goods and services from Sweden to foreign countries. This will increase (decrease) aggregate demand in the small export-dependent Swedish economy, and therefore affect the Swedish economy in a positive (negative) way.

One potential issue with our manual sentiment classification is that it may deviate from the general public's perception of economic media sentiment in a systematic way. This is not likely to cause any substantial problems in our analysis, however. Many experimental studies that investigate to what extent individuals agree on sentiment in texts find an inter-annotator agreement rate in the range of 80 % – 90 % (*e.g.* Wilson *et al.* 2005; Bermingham & Smeaton 2010; Jain & Nemade 2010). The inter-annotator agreement rate is even higher when annotators are required to report on texts' sentiment at a low-granular level, *i.e.*, when annotators are not given any option to express their opinion of the strength of a text's sentiment. The highest agreement among annotators, *i.e.*, when all annotators agree completely, is achieved when an experiment requires annotators to conduct binary classification, *i.e.*, to classify texts as positive or negative sentiment without a neutral option (O'Hare *et al.* 2009).

**3.2.2 Prediction Model**

We use our training data to build a prediction model using the multinomial Naïve Bayes classifier (NB). NB is a robust and parsimonious SML-technique that assigns a class to a given text document by finding the maximum *a posteriori* class under the assumption that each word occurs independently of all other words in a language. In practice, this means that NB assigns a document, which is *not* in the training data, to the class which has the largest sum of log prior probability and log conditional probabilities across words in the documents. NB calculates the prior probability for a document belonging to a particular class as the relative frequency of documents that belong to the class *in* the training data. The conditional probabilities for words in a class are calculated as the relative frequency of words in documents belonging to the class *in* the training data. The NB classifier computes the probability of a class *c*, given a document *d*, as proportional to the product of the prior probability of a document class *P(c)* and the probabilities of each word, *w*, of which there are *n* in a document, *d*, given *c*:

$$P(c|d) \propto P(c) \prod_{1 \leq k \leq n_d} P(w_k|c) \qquad (1)$$

Here, $P(w_k|c)$ is a measure of how much evidence $w_k$ contributes to the class, *c*, being correct for the document, *d*. We are interested in finding the maximum *a posteriori* class, $c_{map}$, for a given document. This means that we can formulate the classification problem as:

$$c_{map} = \arg\max_{c \in C} P(c|d) = \arg\max_{c \in C} P(d|c)P(c)/P(d) = \arg\max_{c \in C} P(d|c)P(c) \qquad (2)$$

Where the first equality is due to Bayes' Rule, and where the second equality comes from the fact that *P(d)*, the probability of a document, is the same across all documents. Here, *P(d|c)* is the probability of a document, conditional on a class, *c*. Estimating this probability is not feasible in practice because it requires that one observes each possible sequence of words for each class, which implies that one needs a near-infinite number of documents in the test data. We thus make the conditional independence assumption, *i.e.* that sequences of words are independent across classes. Additionally, we make a positional independence assumption, *i.e.* that the probability of a word, *w*, conditional on a document, *d*, is independent of the words' position, *k*, in the sequence of words which constitute a document. This assumption is clearly *naïve* because it assumes that words occur in random sequences in language. The conditional- and positional independence assumptions allow us to rewrite (2) as:

$$c_{map} = \arg\max_{c \in C} P(c) \prod_{1 \leq k \leq n_d} P(w_k|c) \tag{3}$$

Here, we take logs to avoid numerical underflow when we estimate the model in practice:

$$c_{map} = \arg\max_{c \in C} \log(P(c)) + \sum_{1 \leq k \leq n_d} \log(P(w_k|c)) \tag{4}$$

We can now obtain estimators for *P(c)* and *P(w_k|c)* by using maximum likelihood which renders the following closed form solutions:

$$\widehat{P}(c) = \frac{N_c}{N} \tag{5}$$

$$\widehat{P}(w_k|c) = \frac{W_{cw}}{\sum_{w \in V} W_{cw'}} \tag{6}$$

Where *N* is the total number of documents in the training data, $N_c$ is the number of documents in the training data that belongs to class *c*, $W_{cw}$ is the word count for word *w*, in all documents in the training data that belong to class, *c*. The sum, $\sum_{t \in V} W_{cw}$, adds the counts of all words in the vocabulary *V* for documents that belong to class *c*, in the training data. The vocabulary is the list of all unique words in all of the documents in the training data. In practice, equation 3 may not be defined because we may have instances of word counts which are zero, *i.e.* $W_{cw} = 0$. We solve this issue by updating equation

(5) with Laplace smoothing, also known as add-one smoothing, which is a standard approach in the literature:

$$\widehat{P}(w_k|c) = \frac{W_{cw}+1}{\sum_{t \in V}(W_{cw'}+1)} \tag{7}$$

We estimate (5) and (7) in the training phase using our training data. These estimates are then used in equation (4) to classify documents which are not in the training data.[5]

NB performs particularly well in terms of prediction compared to other SML-techniques, such as logistic regression, when the training data is small in relation to the size of the feature space (Ng & Jordan 2002). This makes NB well suited in our case since our feature space consists of 33,660 unique words stemming from the texts we use as training data. We use 50-fold cross-validation and find that our NB-model classifies media items into the classes (i)–(iii) with an estimated overall accuracy of 70 %. This is a good prediction rate in relation to other studies that classify texts on sentiment, which in general report average prediction rates ranging from 52 % - 64 % (Serrano-Guerrero *et al.* 2015; Boydstun *et al.* 2018). Our NB-model is likely to classify a document containing words like *"strong"*, *"good"* and *"growth"* as positive sentiment rather than negative sentiment. Our NB-model is, however, likely to classify a document containing words like *"weaker"*, *"reduced"* and *"recession"* as negative sentiment rather than positive sentiment.[6]

### 3.2.3 Classified Data

We use our NB-model to classify the full sample of media items which results in 31% economic media items classified as positive sentiment, 24 % classified as negative sentiment and 45 % classified as irrelevant. The ratio of positive media items in the full data set is the same as in the test data. The ratios of negative and irrelevant data are, however, different. The test data contained 5 percentage points fewer irrelevant media items and 5 percentage points more negative media items. This difference is most likely due to a random overrepresentation of the negative class in the test data sample. Figure 2 shows the number of news items for the 20 most frequent media outlets in the relevant data. *Dagens Industri* (DI), which is the Swedish equivalent to the *Financial Times*, is the most frequent media outlet in the data with 10% of the news items. The second most common outlet is *Dagens Nyheter* (DN) with 6% of the items in the data, followed by *Affärsvärlden* (AFV) with 5% of the items in the data, *Svenska Dagbladet* (SD) with 5% and *Webfinanser* (WF) with 4% of the media items in the data respectively. Out of these five, the most frequent outlets DI, DN and SD are daily newspapers with nationwide coverage, AFV is a weekly newspaper with nationwide coverage and WF is an online news platform. It should be noted that both digital and printed media items are included in the counts presented in Figure 2.

**Figure 2. Top-20 Most Frequent Media Outlets**

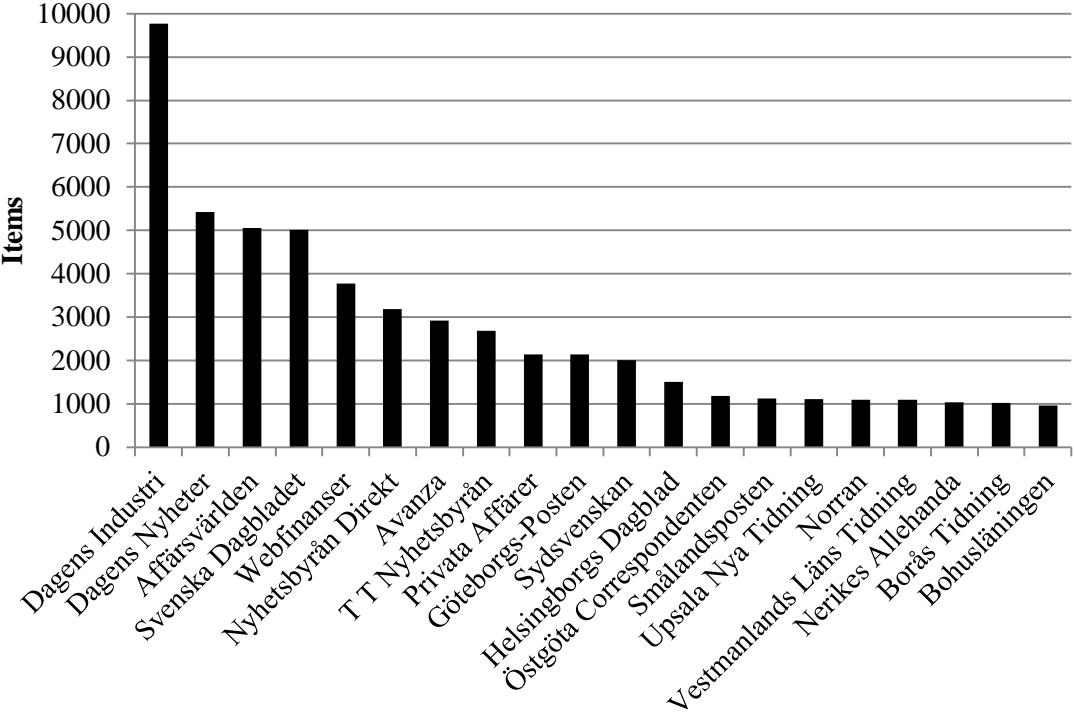

The top five outlets in the data are followed by a set of newspapers and websites with nationwide coverage, all of them with more than 1,500 news items in the data. The news outlets with around 1,000 news items in the data are all local newspapers. The average number of words in a media item for the top 20 media outlets is between 284.5 and 775.9, covering the bulk of the empirical distribution as presented in Figure 3. Most media items are short. The average number of words in an item is 569.

There is, however, a wide variation in the average length of the outlets' items, as illustrated in Figure 3. Most of the outlets in the data publish text items that are shorter than 1,000 words on average. This indicates that most of the media outlets in the data are news agencies that frequently publish short stories about economic developments. The media outlets that on average publish the longest texts (>1,000 words) are different from newspapers as they publish in-depth analyses, forecasts and minutes from board meetings, *etc.* at a low frequency.

**Figure 3. Media Outlets' Average Number of Words per News Item**

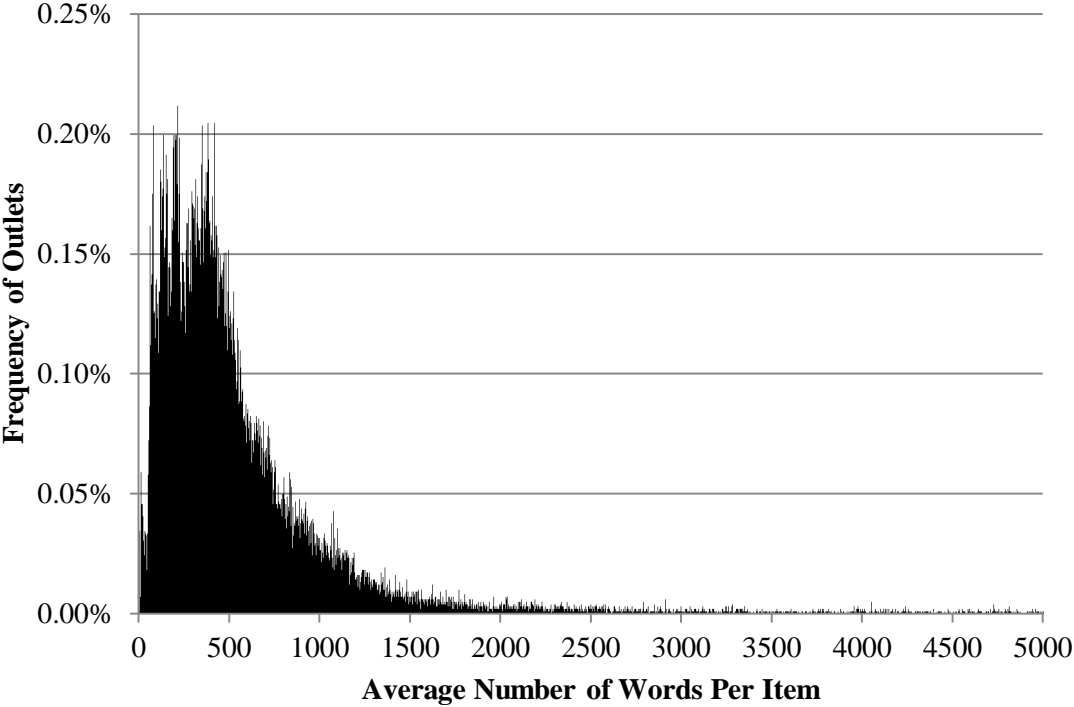

In our analysis, we split the media items into six different subgroups in order to investigate if they affect individuals' opinions and expectations about economic conditions in different ways. The first split we make concerns publishing format, *i.e.*, whether media items are published in print or online. Printed media will probably follow stricter editorial procedures than online media, which are published more frequently than printed media. Their sentiment may differ from online media, which are published continuously throughout the day, therefore not allowing as strict an editorial process as can be expected in printed media. The second distinction we make concerns geographical affiliation, *i.e.*, between media items published by media outlets which have nationwide coverage and those with only local coverage. We define nationwide coverage as media outlets that do not have an explicit local affiliation in their name or in their description of themselves as stated on their homepage. This split allows us to identify differences in media sentiment that stem from the local media, which is more likely to be more concerned with local economic conditions, and nationwide economic media sentiment, which is more likely to report nationwide economic conditions. The local media may thus affect individuals' opinions and expectations about their own personal economic situation, since local economic conditions are more likely to be closely linked to individuals' personal economic conditions than nationwide economic conditions. The third distinction we make is based on two features of the media outlets that can be found in the data. The first feature is the media outlets' publication frequency, *i.e.*, how many media items they have in the data. The second feature is the media outlets' average number of words per media item in the data[7]. This will allow us to investigate the differences between media items that publish in-depth analysis, *i.e.*, that are published infrequently and are relatively long, and media items that are published frequently with a relatively shallow content.

### 3.3 The Economic Media Sentiment Index

We calculate an Economic Media Sentiment Index (EMSI) by taking monthly averages of the ratio of the net of positive and negative media items to total number of economic media items for each month, according to the following formula.

$$EMSI_t = D_t^{-1} \sum_{d=1}^{D_t} net_d \tag{8}$$

Where *d* is day in the month *t* with $D_t$ number of days in the month, and where:

$$net_d = \begin{cases} \dfrac{\#\,positive_d - \#\,negative_d}{\#\,positive_d + \#\,negative_d}, & \text{if } \#\,positive_d + \#\,negative_d > 0 \\ 0, & \text{if } \#\,positive_d + \#\,negative_d = 0 \end{cases} \tag{9}$$

This gives us an index which ranges from -1 when there is only negative news during the month, to +1, when there is only positive news. In our data, the index has a mean value of 0.06 with a standard deviation of 0.18. Figure 4 shows the development of the economic media sentiment index for the sample period. The EMSI is relatively volatile with monthly variations of +/- 50% being common. The index also contains some longer cycles, which are illustrated in the figure by using a 12-month centred moving average. In general, the longer cycle follows the business cycle, with peaks around the dot-com boom in 2000, before the international financial crisis in 2008 and during the recovery from the financial crisis in 2010. We can observe the largest troughs in the index during the international financial crisis of 2008/09 and the Euro-zone debt crisis in 2012/13. Other troughs in the EMSI are small in comparison with the financial and Euro crises, for example the trough following the dot-com boom and the one following the Swedish financial crisis in the early 1990s. In fact, the amplitude of the cycles appears to have increased over time. This may be due to more volatile economic conditions and/or to changed media behaviour where media outlets report economic events with stronger sentiment. Such a change in media behaviour may be the result of including media items published online after the year 2000, which do not necessarily follow the same editorial procedures as traditional printed media.

**Figure 4. EMSI and 12-Month Centred Moving Average of EMSI 1993–2017**

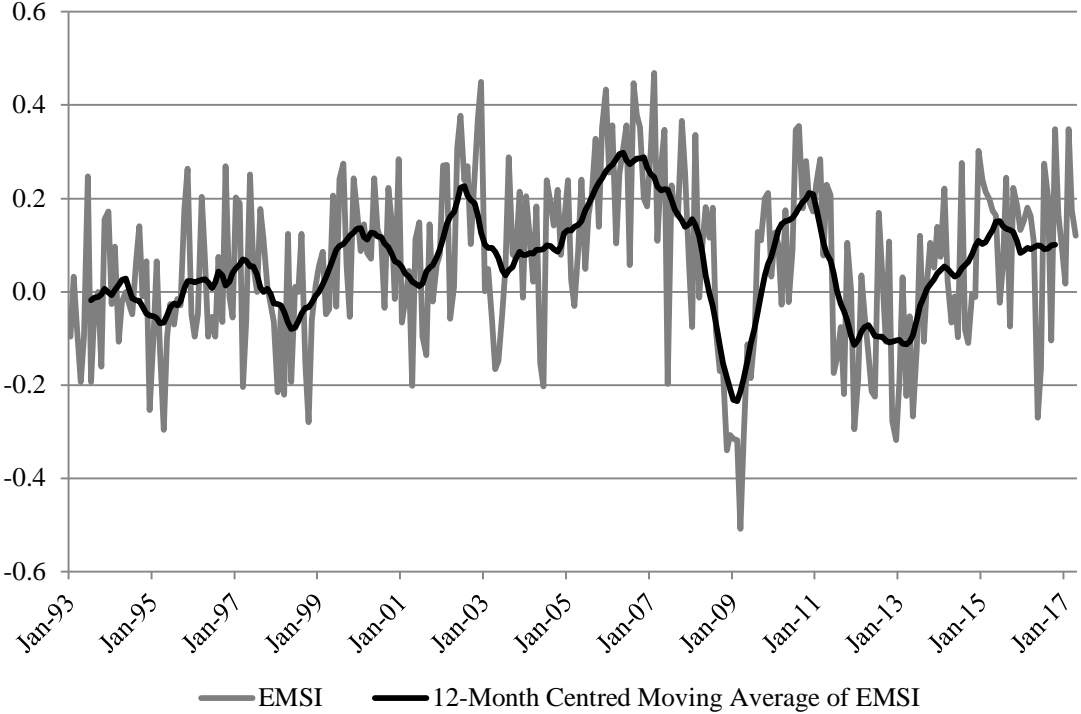

Figure 5 shows the development of the 12-month centred moving averages of the printed media and the online media, together with the output gap as calculated by using Swedish quarterly real GDP from the St. Louis Federal Reserve Economic Data (FRED) database (we have interpolated to obtain monthly observations and used the HP-filter to estimate the output gap). Printed and online EMSI follow the development of the output gap relatively closely throughout the sample, with troughs and peaks occurring around the same time as for the output gap. The output gap shows clear signs of increased volatility from 2006 onwards. Most of this volatility is due to the financial crisis. If we disregard the 2006–2011 period, which covers and is adjacent to the financial crisis, we do not observe any signs of increased volatility in the output gap. The online EMSI is, however, more positive than the printed EMSI particularly during the 2000–2005 and 2011–2017 periods. The increased variance of total EMSI may thus come from both increased volatility in the economy during the 2006–2012-period and also from the influx of online media in the data. Another observation is that both online and printed EMSI appear to lead the development in the output gap, particularly for the 2006–2011 time periods. Both print and online EMSI reach a pre-financial-crisis peak prior to 2007, while the output gap peaks after 2007. The EMSI measures and the output gap hit their lowest level at the end of 2008, and the EMSI measures recover more quickly from the trough than the output gap. This pattern, with EMSI reaching peaks and troughs before the output gap, is present throughout the sample but is more pronounced after the year 2000, and is stronger for online EMSI than for print EMSI. This may be due to changed media behaviour, where the media reports more about forecasts than previously.

**Figure 5. 12-month Centred Moving Averages of Printed and Online EMSI and the Output Gap**

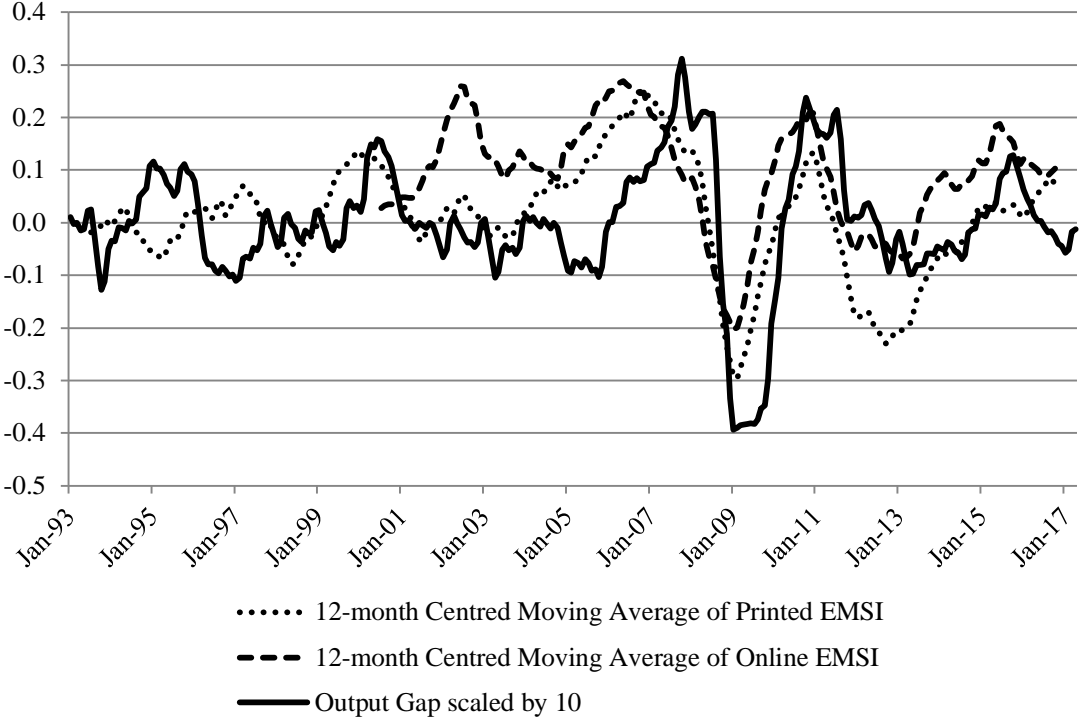

The different subgroups of the EMSI develop similarly to total EMSI during the sample period. Figure 6 shows that the development of nationwide, local, printed, online, frequent and infrequent EMSI develop in cyclical movements that are similar to those for total EMSI. The volatility of the subgroups are different, however. Printed media appears to be more volatile than nationwide media. This may be because nationwide media are better funded than other media outlets that are published in print, and may thus present a more balanced economic sentiment in their media items. The subgroups of EMSI (local, online and infrequent), which enter into the data in the year 2000, are more volatile at the beginning of the 2000s than later in the sample period. This may be because the number of online media outlets in the Retriever database has gradually increased from the year 2000 and onwards as the importance of online media increased in the media landscape. This should have a stabilizing effect on our EMSI measure since the larger number of online media outlets makes the EMSI less sensitive to the sentiment of merely a few online media outlets.

**Figure 6. Subgroups of EMSI 1993–2017**

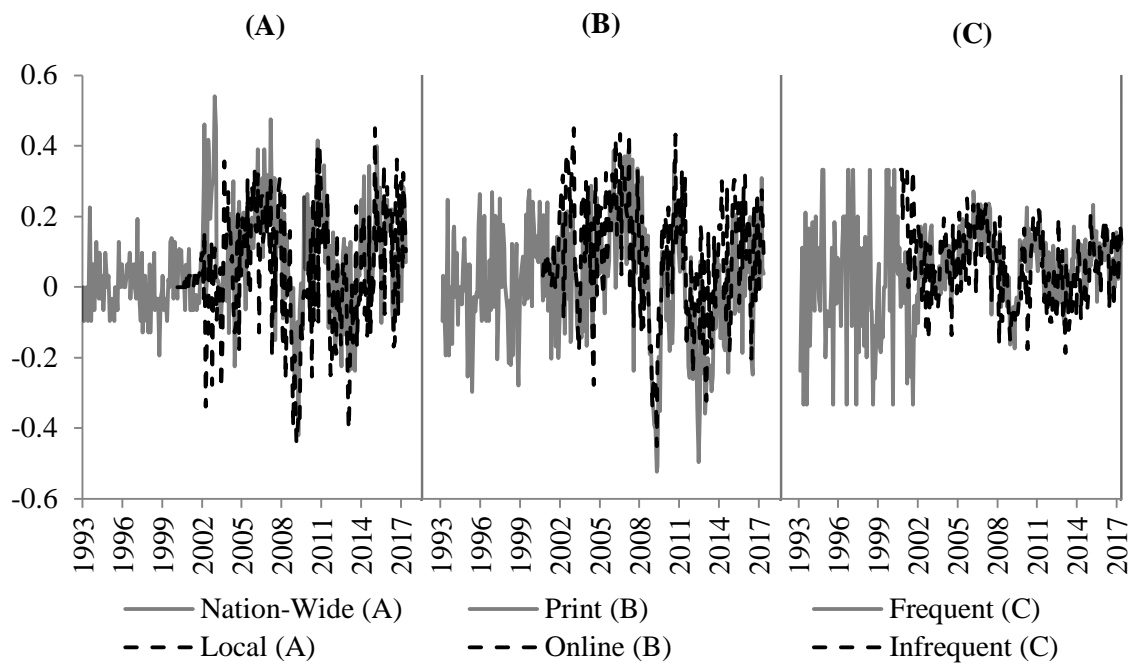

## 4. Data on Individuals' Opinions and Expectations

Having quantified the economic media data, we now turn to the measurement of individuals' evaluations and expectations of economic conditions. Researchers often measure these by using survey data (*e.g.* Boydstun *et al*. 2018; Fogarty 2005; Hester & Gibson 2003). We use data on individuals' opinions and expectations about economic conditions that are collected in the Swedish Household Purchasing Plan's survey (*Hushållens inköpsplaner*, henceforth referred to as "the survey"), which is a monthly survey conducted by the National Institute of Economic Research (NIER).[8] The survey is representative of the adult population and each wave includes 1,500 respondents.[9].The survey contains more than 16 questions about the household's opinions concerning their current and future consumption and their opinions about current and future economic conditions. Van Raaij and Gianotten (1990) have shown that questions in consumer sentiment surveys can be divided into two main groups: questions that represent the individual's sentiment about their own economic conditions and questions that represent the individual's sentiment about the nation's overall (macro) economic conditions. In order to make our analysis as general as possible, we have thus chosen to use four questions in the survey that are likely to best represent these two groups. These questions are presented in Table 1.

**Table 1. Survey Questions and Answers**

| Name | Question Asked to Respondents in the Survey |
|---|---|
| ownFuture | "How do you think the fiscal position of your household will change over the next 12 months?" |
| sweFuture | "How do you think the general economic situation in this country will develop over the next 12 months?" |
| ownNow | "How is the fiscal position of your household today compared with 12 months ago?" |
| sweNow | "How is the general economic situation in this country compared to 12 months ago?" |
| | **Alternative Answers offered to Respondents in the Survey** |
| A1 | "Got/Get a lot worse in relation to 12 months ago/today" |
| A2 | "Got/Get a little worse in relation to 12 months ago/today" |
| A3 | "The same in relation to 12 months ago/today" |
| A4 | "Got/Get a little better in relation to 12 months ago/today" |
| A5 | "Got/Get a lot better in relation to 12 months ago/today" |
| A6 | "I don't know" |

One set of survey questions asks individuals to state their expectations about the development of their own private economic situation and the macroeconomic situation during the upcoming 12 months (ownFuture and sweFuture). Another set of questions asks respondents how they evaluate the development of their own economic situation and the macroeconomic situation during the past 12 months (ownNow, sweNow). Using the answers to these questions allows us to evaluate if an individual's expectations and evaluations of past economic developments are caused by the EMSI. Respondents can give one out of six predefined answers to the questions. These are presented in Table 1, and range from very negative (A1) to very positive (A5), including a neutral option (A3) and the option to report a lack of knowledge (A6). In our econometric models, we use the average of within month-waves of the net of positive and negative replies to these questions, which are calculated as the sum of the number of positive answers (A5, A4) minus the sum of the number of negative answers (A1, A2) within each month.

## 4.1 Macroeconomic Data

We need to control for macroeconomic variables in our analysis in order to isolate the sentiment that the news media adds when reporting economic conditions. We therefore include a set of key macroeconomic variables in our regression models. These macroeconomic variables are selected in order to capture the most important variations in the business cycle, the financial markets and exogenous shocks that hit the economy at random.

We use quarterly real GDP which we download from the St. Louis FRED database as a measure of the business cycle in terms of economic activity. We interpolate this variable to obtain monthly observations and estimate the output gap using the HP-filter, which is a very common approach in the literature (Ravn & Uhlig 2002). Interpolating the GDP data may introduce some measurement error in the data (Friedman 1962). However, this measurement error is not likely to be a major problem in our analysis since GDP is a variable that changes slowly with relatively little variation between quarters. The variation between months is thus also likely to be relatively small and thus the measurement error as well. Furthermore, the media items in our data report findings from reports and forecasts published by economic bodies. Some of these reports and forecasts use models that also include past values of interpolated GDP from quarterly to monthly frequency, together with other key economic indicators available at monthly frequency, to forecast current and future GDP (*e.g.* Mitchell *et al.* 2005; Altissimo *et al.* 2010). Using the monthly interpolated GDP in our analysis should therefore not deviate systematically from what is being done in these reports and forecasts. This is especially so since we furthermore use monthly data available from statistics Sweden (SCB) for the unemployment rate (to account for business cycle fluctuations), and for the Consumer Price Index (CPI) to calculate the yearly inflation rate.

We include two measures of exogenous shocks in our model. Firstly, we use monthly oil prices from the World Bank, which we convert to year-on-year percentage changes in real values, and secondly the monthly SEK/EUR-exchange rate from the Swedish Riksbank. Both of these variables may be important for explaining individuals' expectations and perceptions of economic circumstances as well as being important for the supply side of the economy. Individuals tend to attach a symbolic value to the national exchange rate, which may affect their perception of the performance of the macro economy (Hobolt & Leblond 2009). The oil price directly affects individuals' personal economic situation through the price of petrol used for private transport (Johnson & Lamdin 2012).

We use the monthly stock prices in terms of the OMXS30 index from Nasdaq OMX Nordic, which are transformed to real values using the CPI and then converted to year-on-year percentage changes. Stock prices are likely to matter both for individuals' perceptions of their own personal economic situation and the nationwide economic situation. Swedish households' participation in the stock market is high, with 45% of households owning stocks either directly as a part of their savings portfolio or indirectly through shares in mutual funds (Almenberg & Dreber 2015). Stock prices may also matter for individuals' perceptions and expectations concerning the future of the macro economy, since stock prices tend to lead the development of the real economy (Fama 1990; Schwert 1990). We henceforth prefix all macroeconomic variables that we calculate using year-on-year percentage change with ∆%.

## 4.2 Descriptive Statistics

Table 2 shows descriptive statistics for the variables that we include in our econometric models. EMSI ranges from -0.51 to 0.47, which means that it is at most approximately half of the extreme values (-1 and 1) for this measure, *i.e.*, when the news media reports only negative or positive news items during all the days of the month. We further note that ownFuture, sweFuture and ownNow have positive averages, which indicates that individuals are, on average, optimistic about the state of their own present and future economic situation and about the future of the nationwide economic situation. However, the average net of replies to sweNow is negative, which indicates that individuals are, on average, pessimistic when evaluating the state of the present nationwide economic situation.

The survey measures of individuals' own economic situation are more stable than the survey measures concerning individuals' evaluations and expectations about the future of the nationwide economy. The standard deviations of sweNow and sweFuture are more than twice as large as the standard deviations for ownNow and ownFuture. This indicates that individuals base their expectations and evaluations about the state of their own personal economic situation and the nationwide economic situation on information sets that are not identical. The information set used by individuals to form an opinion about their own personal economic situation is likely to be more precise and less volatile compared to the information used by individuals to form their opinions about the nationwide economy. It is thus likely that the media plays an important role for individuals' expectations and opinions about the nationwide economy given the rather choppy development of EMSI presented in Figure 1. The p-values from the Augmented Dickey-Fuller test including a constant shows that we reject the null of a unit root at a 5%-level for all of our variables except the inflation rate. Although inflation may be nonstationary in theory, it is counterintuitive that the inflation rate is nonstationary as it would imply, for example, infinite hyperinflation. A plausible explanation for this result is that the Swedish central bank (*Riksbanken*) was given an independent mandate to conduct monetary policy with an inflation target of 2% in 1995. This is likely to have caused the inflation rate to move from a higher level of stationary inflation to a lower level after 1995. In our analysis, we thus use an inflation rate which is demeaned with the pre- and post-1995 means for the respective periods. This inflation rate has an ADF-test p-value of 0.045.

**Table 2. Descriptive Statistics**

|  | Mean | Median | Minimum | Maximum | Std. Dev. | ADF-test P-value |
|---|---|---|---|---|---|---|
| **EMSI** | 0.06 | 0.07 | -0.51 | 0.47 | 0.18 | 0.00 |
| **ownNow** | 4.88 | 8.5 | -34.9 | 19.5 | 11.68 | 0.02 |
| **ownFuture** | 16.29 | 18.4 | -28.6 | 32.2 | 9.02 | 0.00 |
| **sweNow** | -12.16 | -11.6 | -92.8 | 42.3 | 27.47 | 0.00 |
| **sweFuture** | 6.9 | 5.2 | -43.2 | 49.9 | 19.98 | 0.01 |
| **Output Gap** | 0.00 | 0.00 | -0.04 | 0.03 | 0.01 | 0.00 |
| **Unemployment Rate** | 7.60 | 7.65 | 4.94 | 10.48 | 1.25 | 0.03 |
| **Inflation Rate** | 1.31 | 1.14 | -1.55 | 5.09 | 1.35 | 0.12 |
| **SEK/EUR** | 9.14 | 9.14 | 8.24 | 11.17 | 0.49 | 0.04 |
| **Δ%Oil Price** | 8.54 | 2.05 | -56.46 | 168.00 | 36.68 | 0.01 |
| **Δ%Stock Prices** | 8.03 | 13.10 | -77.38 | 96.34 | 31.66 | 0.00 |

The p-values are from the Augmented Dickey-Fuller test including a constant. The ADF-test p-value for the inflation rate is 0.045 when we level-adjust it with regards to a change in monetary policy regime in 1995. This is the version of the inflation rate that we use in our analysis.

The correlations in Table 3 confirm the observed pattern in Figures 3 and 4, *i.e.*, that EMSI is correlated with the business cycle. The correlation between EMSI and the output gap is 0.30 and the correlation between EMSI and unemployment is -0.23. The EMSI is more volatile compared to the macroeconomic variables, which may explain the relatively weak correlations. Correlations based on a 12-month centred moving average of EMSI are more highly correlated with the macroeconomic variables (0.43 with the output gap and -0.35 with unemployment). EMSI is positively correlated with all of the net of individuals' replies to the survey questions. The highest correlation, between EMSI and individuals' evaluation of the current developments of the Swedish economy (sweNow), is 0.44, and is 0.60 for the 12-month centred moving average of EMSI. The strength of the correlation between the one month lag of EMSI and the survey- and macro variables is similar to the strength of the contemporaneous correlations. This indicates that there is a potential between-month causal relationship from the one period lag of EMSI to the contemporaneous survey variables.

**Table 3. Correlations**

|  | Contemporaneous | | One Month Lag | |
| --- | --- | --- | --- | --- |
|  | EMSI | 12-Month Centred Moving Average | EMSI | 12-Month Centred Moving Average |
| **sweNow** | 0.44 | 0.60 | 0.47 | 0.62 |
| **sweFuture** | 0.18 | 0.17 | 0.16 | 0.13 |
| **ownNow** | 0.31 | 0.44 | 0.29 | 0.43 |
| **ownFuture** | 0.30 | 0.45 | 0.28 | 0.44 |
| **Output Gap** | 0.30 | 0.43 | 0.36 | 0.50 |
| **Unemployment Rate** | -0.23 | -0.35 | -0.21 | -0.38 |
| **Inflation Rate** | -0.02 | 0.02 | 0.07 | 0.10 |
| **SEK/EUR** | -0.08 | -0.06 | -0.12 | -0.09 |
| **Δ%Oil Price** | 0.25 | 0.33 | 0.28 | 0.35 |
| **Δ%Stock Prices** | 0.20 | 0.24 | 0.18 | 0.21 |

## 5. Results

### 5.1 Test of Hypothesis 1

We can now test Hypothesis 1, that economic media is not neutral and therefore publishes unequally many items with positive and negative sentiment. As our data consists of monthly observations spanning 24 years, it includes several booms and busts in the real economy. We thus expect a neutral media to publish an equal number of media items with positive and negative tone and a non-neutral media to publish an unequal number of media items with positive and negative tone. We thus test Hypothesis 1 by statistically testing if the mean of EMSI is significantly different from zero. Table 4 shows the sample means and t-ratios calculated using OLS with the EMSI as dependent variable and a constant as independent variable, using robust standard errors (Newey & West 1987). Table 4 also includes tests for our six different subgroups of EMSI including sources with nationwide coverage, local coverage, online and printed modes of publication and frequent and infrequent publications. We can see that the total EMSI, together with five out of the six subgroups, are positive and significantly different from zero, causing us to accept Hypothesis 1. The interpretation of the estimated mean values is straightforward: a positive and significant mean value means that the number of positive media items is on average greater than the number of negative ones. The estimated mean for total EMSI is 0.06, which indicated that in a month with 300 economic media items there will be 18 more items during the month with a positive tone than with a negative tone. The estimated mean values of EMSI and the subgroups are thus only *moderately* positive on average.

**Table 4. Significance Test of Means**

| EMSI | Nationwide | Print | Local | Online | Frequent | Infrequent |
|---|---|---|---|---|---|---|
| 0.06*** | 0.07*** | 0.01 | 0.04** | 0.09*** | 0.04*** | 0.01*** |
| (3.62) | (4.94) | (0.69) | (2.21) | (4.72) | (4.20) | (5.10) |

*= p-value<0.1, **= p-value<0.05, ***= p-value<0.01, t-ratios presented in parentheses. The means are estimated using OLS with the EMSI as dependent variable and a constant as an independent variable. The regressions are conducted using robust standard errors (HAC).

Next, we investigate the difference in means and variances between the subgroups presented in Table 5. Firstly, we can note that economic news items that are published online have a predominately more positive tone than those published in print. This difference could be due to the fact that traditional printed media employs a stricter editorial process and is published less frequently than online news media, causing the traditional media to present a more neutral version of economic conditions on average. We further note that the nationwide economic media is more positive than the local media. This could be due to the local media mainly re-reporting macroeconomic news from the nationwide media, to which it adds local economic news with a more negative tone. We find no significant difference between the variances of the different subgroups. This indicates that the sentiment of the subgroups reacts to changes in economic conditions in a similar way, albeit with different averages.

**Table 5. Significance Tests for Differences in Means and Variances**

| EMSI-measures | Difference in mean | Difference in variance |
|---|---|---|
| **Print – Online** | -0.08*** <br> (−5.30) | 0.01 <br> (1.14) |
| **Nationwide – Local** | 0.05*** <br> (3.70) | 0.00 <br> (1.21) |
| **Shallow – Deep** | -0.00 <br> (−0.33) | 0.00 <br> (1.23) |

*= p-value<0.1, **= p-value<0.05, ***= p-value<0.01. The differences in means are estimated OLS with the difference in EMSI as dependent variable and a constant as independent variable. The regressions are conducted using robust standard errors (HAC). t-ratios are presented in parentheses for test of differences in means, and F-statistics for tests of differences in variances.

### 5.2 Tests of Hypotheses 2 and 3

We now turn to tests of Hypothesis 2, that individuals' opinions and expectations about current and future economic conditions are Granger-caused by economic media content that is not motivated by economic data, and Hypothesis 3, that economic media content does not Granger-cause individuals' opinions and expectations about their private economic situation. As both of these hypotheses regard causation of economic media on individuals' expectations and evaluations of economic circumstances, we evaluate them using a set of Granger-causality tests, see Granger (1969).

#### 5.2.1 Granger-causality Tests

We acknowledge the fact that Granger-causality is not a test of true causality because it can only establish a causal relationship between two variables, **X** and **Y**, in the sense that one variable tends to occur before the other one in time. Thus, the Granger-causality test cannot confirm that there is a true causal relationship between **X** and **Y** but only rule out that there is no true causal relationship between them. We furthermore acknowledge that the Granger-causality test (i) does not rule out the possibility that two variables **X** and **Y** are caused by a third potentially unobserved variable **Z** and (ii) it does not account for instantaneous causality between **X** and **Y** (Maziarz 2015). Both the EMSI and the general public are likely to respond to changes in economic data. Any one-way Granger-causality established between EMSI and individuals' expectations and evaluations of economic conditions may thus be due to macroeconomic data and therefore not a true causal relationship. This is not likely to cause any problems in our analysis because we include key macroeconomic variables covering the most important sectors of the macro economy as controls in our regressions. It is thus likely that any Granger-causality we can establish between our variables of interest does not come from an unobserved variable that we do not capture with the macroeconomic variables. We test if a causal relationship exists between our survey measures and EMSI using the following specification:

$$survey_t = \alpha + \sum_{k=1}^{k=K} \beta_{0,k} \times EMSI_{t-k} + \sum_{i=1}^{i=6}\sum_{k=1}^{k=K} \beta_{i,k} \times macrovariable_{i,t-k} + \sum_{k=1}^{k=K} \beta_{7,k} \times survey_{t-k} + \varepsilon_t \quad (10)$$

$$EMSI_t = \alpha + \sum_{k=1}^{k=K} \beta_{0,k} \times survey_{t-k} + \sum_{i=1}^{i=6}\sum_{k=1}^{k=K} \beta_{i,k} \times macrovariable_{i,t-k} + \sum_{k=1}^{k=K} \beta_{7,k} \times EMSI_{t-k} + \varepsilon_t \quad (11)$$

We determine the number of lags, $K$, in equations 10 and 11 individually by selecting the most parsimonious specification that does not exhibit autocorrelation in the residuals. If the coefficient for EMSI is significant on at least a 5%-level in equation 10, then, we can conclude that EMSI Granger causes individuals' opinions and expectations about economic conditions, rather than this coming from our economic control variables. We can determine the direction of causation by investigating the significance of the coefficient estimates for our survey variables in equation 8. If all of them are insignificant, then we can conclude that there is a one-way Granger causal relationship going from EMSI to our survey variables, we denote this result as *EMSI → survey*.

If, however, the coefficients in equation 8 are also significant, then we can conclude that there is a two-way Granger causal relationship between EMSI and our survey variables, we denote this result *EMSI ↔ survey*. If we observe that the coefficient for EMSI is insignificant in equation 10 and that the coefficient for our survey variables is significant in equation 8 then we can conclude that there is a one-way Granger causal relationship going from our survey variables to EMSI. This may be due to the media reporting on recent results from the survey as part of, for example, the NIER's Economic

Tendency Survey (*Konjunturbarometern*). We must thus rule out the possibility that our survey measures cause the EMSI and that we thus have a feedback loop between the media and the general public. We denote this result as *survey → EMSI*.

**Table 6. Results from Granger-causality Tests**

|            | sweNow | sweFuture | ownFuture | ownNow |
|------------|--------|-----------|-----------|--------|
| **EMSI**       | →      | ←         |           |        |
| **Online**     | →      |           | ←         |        |
| **Nationwide** | →      |           |           |        |
| **Print**      | →      |           |           |        |
| **Frequent**   |        |           |           | ←      |
| **Local**      |        |           |           |        |
| **Infrequent** |        |           |           |        |

The symbols show Granger-causality between the row variable on the left side of the symbol and column variable on the right side of the symbol, *e.g.* the top left symbol represents EMSI → sweNow.

The results from our Granger-causality tests presented in Table 6 show a clear pattern. Four out of seven tests result in one-way Granger-causality between EMSI and sweNow and three out of seven tests result in no Granger-causality between EMSI and sweNow. These results give support to Hypothesis 2 because the news media Granger-causes individuals' opinions about the macro economy in terms of sweNow, *i.e.* that *EMSI → sweNow*, *Online → sweNow*, *Nation-Wide → sweNow* and *Print → sweNow*. The results also show that there is a one-way Granger-causal relationship going in the other direction from the survey variables to EMSI in terms of *sweFuture → EMSI*, *ownFuture → Online* and *ownNow → Frequent*. These results are, however not robust across different subgroups of EMSI. Hypothesis 3 is supported by the results presented in Table 6 since EMSI does not Granger-cause ownFuture and ownNow. This result indicates that individuals do not base their opinions and expectations about their own economic situation on information published by the economic media.

Individuals' opinions about the current state of economic conditions are Granger-caused by EMSI and Δ% Stock prices. The results for equation 10 with sweNow as dependent variable are presented in Table 7. These show that EMSI is significant for 4 out of 7 specifications with estimates in the range of 5.39–8.57. This implies that the long run effect of EMSI on sweNow is in the range of 59.89–85.7.

The long run effect of EMSI on sweNow in a month with a large share of positive news, such as the sample maximum of 0.47, is thus in the range of 28.15–40.28. This roughly corresponds to an interval ranging from a 1–1.5 standard deviations increase in the net number of positive replies in sweNow. The estimated coefficient for Δ%Stock Prices is positive and significant in all of our specifications of equation 10, indicating that individuals use the stock market as an indicator of current macroeconomic conditions. Lags of the other macroeconomic variables are insignificant in most of the specifications, with the exception of inflation which is significant at the 10% level in the model specifications with EMSI and Print as independent variables.

**Table 7. OLS Regression Results from Equation Specification 7**

| Dependent variable | sweNow | sweNow | sweNow | sweNow | sweNow | sweNow | sweNow |
|---|---|---|---|---|---|---|---|
| Economic Media Sentiment | EMSI | Online | Nationwide | Infrequent | Print | Frequent | Local |
| α | -8.78 (12.26) | -8.14 (13.4) | -7.10 (12.37) | -11.24 (13.87) | -9.10 (12.56) | -8.00 (14.16) | -7.94 (14.51) |
| $EMSI_{t-1}$ | 5.39** (2.48) | 7.53*** (2.47) | 5.48** (2.48) | 8.57** (4.22) | 5.8* (2.95) | 5.81 (4.43) | 5.04 (3.22) |
| $Δ\%Stock\ Prices_{t-1}$ | 0.06*** (0.02) | 0.14*** (0.02) | 0.06*** (0.02) | 0.14*** (0.03) | 0.06*** (0.02) | 0.13*** (0.03) | 0.13*** (0.03) |
| $inflation_{t-1}$ | -0.96* (0.58) | 0.16 (0.95) | -0.95 (0.58) | 0.09 (1.01) | -1.04* (0.58) | 0.14 (1.01) | 0.17 (1) |
| $unemployment_{t-1}$ | 0.16 (0.48) | -0.02 (0.69) | 0.09 (0.48) | 0.14 (0.71) | 0.11 (0.48) | -0.19 (0.76) | -0.1 (0.73) |
| $Output\ Gap_{t-1}$ | 23.69 (73.1) | -5.08 (80.8) | 19.95 (72.77) | 6.64 (82.74) | 25.09 (73.02) | -11.53 (84.56) | -17.61 (84.92) |
| $SEK/EUR_{t-1}$ | 0.70 (1.2) | 0.64 (1.4) | 0.57 (1.2) | 0.87 (1.48) | 0.81 (1.22) | 0.82 (1.43) | 0.75 (1.49) |
| $Δ\%Oil\ Price_{t-1}$ | -0.01 (0.01) | -0.01 (0.02) | -0.01 (0.01) | -0.01 (0.02) | -0.01 (0.01) | -0.02 (0.02) | -0.02 (0.02) |
| $sweNow_{t-1}$ | 0.91*** (0.03) | 0.90*** (0.03) | 0.92*** (0.03) | 0.90*** (0.03) | 0.91*** (0.03) | 0.91*** (0.03) | 0.91*** (0.03) |
| Adjusted R-sq. | 0.93 | 0.95 | 0.93 | 0.95 | 0.93 | 0.95 | 0.95 |
| Obs. | 291 | 208 | 291 | 208 | 291 | 208 | 208 |

*= p-value<0.10, **= p-value<0.05, ***= p-value<0.01, Robust standard errors (HAC) presented in parentheses.

The estimated coefficients for the one period lagged values of sweNow range from 0.90–0.92. These estimates have 95 % confidence intervals that do not include unity, a result which is further reinforced by the fact that the Augmented Dickey Fuller test rejects the null hypothesis of a unit root for all of our survey measures.[10] The residuals in the regressions presented in Table 7 are not autocorrelated and the results are robust to adding more lags of the dependent variable and the independent variables.

The output gap Granger-causes EMSI, online EMSI and nationwide EMSI in a majority of the regression specifications of equation 8, *i.e.*, with our 7 different EMSI measures as dependent variables and our 4 different survey variables as independent variables using the same macroeconomic control variables in each regression. Figure 7 presents the number of regressions for which lags of the macroeconomic control variables are significant. The lags of the output gap are significant in a majority of the regressions with EMSI, nationwide EMSI and local EMSI as dependent variable. EMSI is also Granger-caused by Δ% Stock prices in 3 out of 4 regressions, which shows that EMSI is Granger-caused both by economic data which changes frequently and infrequently. This indicates that the media are reporting both on economic events that change on a daily basis such as the stock market, as well as changes in other more long-term economic conditions that affect the real economy.

The subgroups of EMSI are Granger-caused by different macroeconomic variables. Online EMSI is Granger-caused by the output gap, while printed EMSI is Granger-caused by measures of exogenous supply-side shocks in terms of the SEK/EUR-exchange rate and Δ% Oil price. The nationwide EMSI is Granger-caused by the output gap, while the local EMSI is Granger-caused by unemployment. This suggests that local media report on topics that are directly linked to individuals' own financial situation to a greater extent than nationwide media, which instead focuses on economic conditions that affect the macroeconomic situation and thus have a weaker direct link to individuals' own economic situation.

**Figure 7. Significant Lags of Macroeconomic Variables in Equation Specification 2 across 4 Different Survey Variables as Independent Variables**

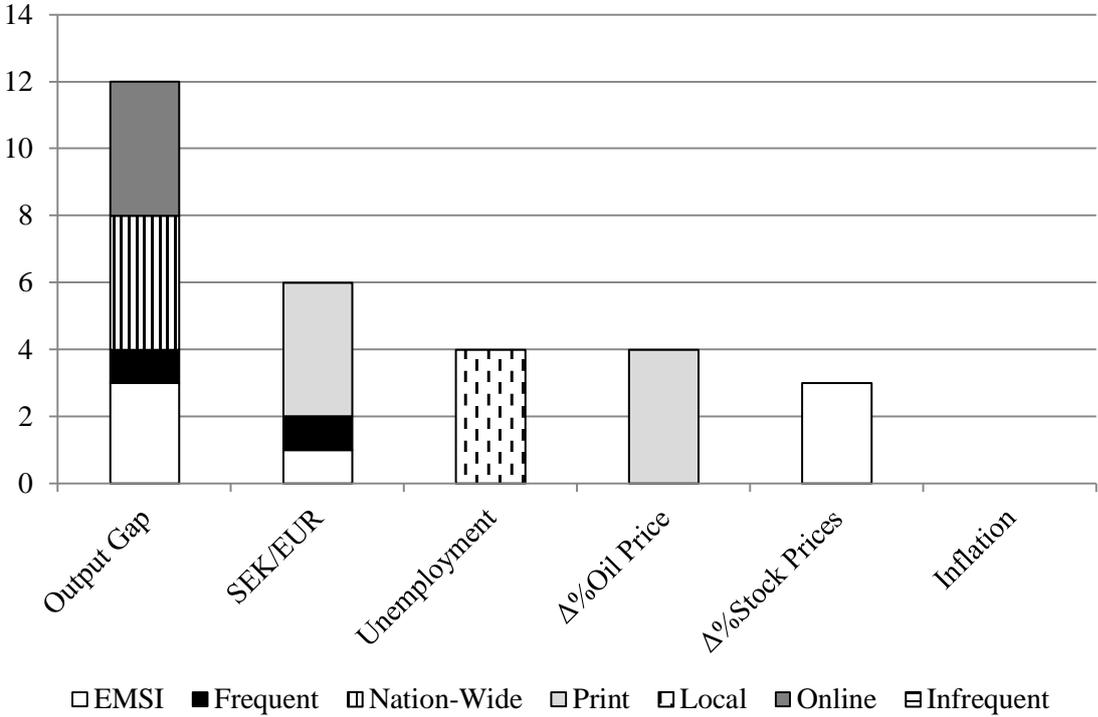

The general picture is that EMSI is Granger-caused by data on economic conditions and not by individuals' expectations and opinions about economic conditions. The results thus indicate that EMSI is Granger-caused by economic conditions and that the EMSI affects individuals' perception of macroeconomic conditions. These are, however, not Granger-caused by economic data apart from stock prices. This indicates that individuals' perception of economic conditions primarily comes from the media.

## 5.2.2 Contemporaneous Effects

We would ideally like to conduct Granger-causality tests using data with a higher frequency than monthly. The macroeconomic data does not allow for this, however, and any attempt at identifying Granger-causality using, for example, daily data will thus suffer from the unobserved variable problem described in section 5.2. We can, however, get an indication of whether a causal relationship exists within a month, albeit without pinpointing the direction of the causality. We do this by running the following regression:

$$survey_t = \alpha + \beta_0 \times EMSI_t + \sum_{i=1}^{6} \beta_i \times macrovariable_t + \sum_{k=1}^{k=K} \beta_{7,k} \times survey_{t-k} + \varepsilon_t \qquad (12)$$

Here, we include the same macro variables as in equations (10) and (11). Although the output gap and the unemployment and inflation rates are not available in real time, they may act as proxies for other, unobserved information about economic conditions that individuals have access to. Including the macro variables thus allows us to isolate the contemporaneous correlation between EMSI and our survey variables. We again determine the number of lags, *K*, individually by selecting the most parsimonious specification that does not exhibit autocorrelation in the residuals. If the coefficient for EMSI is significant in equation 12, then it is an indication that EMSI affects individuals' opinions and expectations about economic conditions, other than that coming from our economic control variables. We cannot say anything about the direction of causality, *i.e.* whether EMSI causes our survey measures or if our survey measures cause EMSI or if they both cause each other. We are, however, able to rule out that EMSI and our survey measures are unrelated within months. Table 8 presents the regression results for model 3:[11]

**Table 8. OLS-Regression Results for Equation 12**

|  | sweNow | sweFuture | ownNow | ownFuture |
|---|---|---|---|---|
| α | 3.67<br>(11.9) | 12.86<br>(10.3) | 4.18<br>(4.24) | 12.89**<br>(5.47) |
| $EMSI_t$ | 9.11***<br>(2.47) | 5.59**<br>(2.32) | 0.02<br>(0.97) | 0.38<br>(1.28) |
| $\Delta\%Stock\ Prices_t$ | 0.05***<br>(0.02) | 0.04**<br>(0.02) | 0.02***<br>(0.01) | 0.02*<br>(0.01) |
| $inflation_t$ | -1.22**<br>(0.53) | 0.55<br>(0.42) | -0.38*<br>(0.21) | -0.23<br>(0.28) |
| $unemployment_t$ | 0.33<br>(0.47) | 0.01<br>(0.42) | -0.49**<br>(0.21) | -0.66**<br>(0.27) |
| $output\ gap_t$ | 9.24<br>(7.37) | -18.42***<br>(4.57) | -0.38<br>(2.44) | -2.78<br>(2.55) |
| $SEK/EUR_t$ | -0.85<br>(1.12) | -1.39<br>(0.99) | 0.01<br>(0.44) | -0.59<br>(0.49) |
| $\Delta\%Oil\ Price_t$ | 0.00<br>(0.02) | -0.03**<br>(0.01) | -0.01**<br>(0.00) | -0.01<br>(0.01) |
| $survey_{t-1}$ | 0.88***<br>(0.03) | 0.92***<br>(0.02) | 0.63***<br>(0.07) | 0.63***<br>(0.06) |
| $survey_{t-2}$ | - | - | 0.22***<br>(0.07) | -0.02<br>(0.08) |
| $survey_{t-3}$ | - | - | 0.07<br>(0.05) | 0.24***<br>(0.06) |
| **Adjusted R-sq.** | 0.93 | 0.88 | 0.94 | 0.86 |
| **Obs.** | 291 | 291 | 291 | 291 |

*= p-value<0.10, **= p-value<0.05, ***= p-value<0.01, Robust standard errors (HAC) presented in parentheses. The Breusch–Godfrey test does not reject the null of no autocorrelation in the residuals in all of the model specifications.

The results in Table 8 give support to Hypothesis 3. The coefficient estimates for EMSI are not significant for ownNow and ownFuture, which means that EMSI is not contemporaneously correlated with ownNow and ownFuture. We can thus rule out that the public's opinions and expectations about their own personal economic situation are affected by EMSI within the same month.

The regression results give some indicative support to Hypothesis 2 since the coefficients of EMSI are positive and significant for sweNow and sweFuture. We are thus able to deduce that there is a Granger-causal relationship between EMSI and sweNow and sweFuture within the same month. The magnitude of the coefficient for EMSI is 63 % larger in the regression with sweNow as dependent variable compared to the regression with sweFuture as dependent variable. This result agrees with the contemporaneous correlations presented in Table 2, which are stronger between EMSI and sweNow than between EMSI and sweFuture. This indicates that individuals may use the economic media to form an opinion about the current state of the Swedish economy to a greater extent than when they form their expectations about its future developments.

The estimated coefficients for the year-on-year percentage change of the stock market index are significant and positive for all the survey variables. This is a reasonable result given that stock prices are a leading indicator for the macro economy and also important for individuals' personal savings. The estimated coefficients for the unemployment rate are significant at a 5-% level for ownNow and ownFuture but not for sweNow and sweFuture. This implies that individuals' perception of their own economic situation correlates with the unemployment rate. This may be due to the fact that individuals' perception of their own economic situation is mostly determined by their employment status; if people expect to become unemployed, then they tend to perceive and expect the future of their personal economic situation to deteriorate.

The estimated coefficients for the inflation rate are negative and significant in the regression with sweNow and ownNow as dependent variables at the 5 % and 10 % level respectively. This shows that individuals tend to become more pessimistic about the current state of economic conditions as inflation increases. The estimates for the year-on-year percentage change of the oil price is negative and significant at the 5 % level for sweFuture and ownNow. This indicates that individuals tend to become more negative if oil price increases, both about their own economic situation and about the nationwide economic situation. This is an expected result since changes in oil prices can be the result of an exogenous supply-side shock, but they also affect expenses for households via the price of petrol used for private transport.

## 6. Conclusion

We develop a new measure on Swedish economic media sentiment using a supervised machine learning technique. Our measure is calculated based on the sentiment of the full textual content of 179,846 media items, stemming from 1,071 unique media outlets including the major nationwide newspapers, local newspapers and online news during the 1993–2017 period. The measure has more good characteristics compared to other studies that leave out important sentiment information or are limited to a few time periods and/or media outlets. We use our measure to test three hypotheses regarding how the media reports on the economy, and how this reporting affects individuals' opinions and expectations. Our results support the second hypothesis; that individuals' opinions and expectations about current and future economic conditions are Granger-caused by economic media content that is not motivated by economic data. There is also support for the third hypothesis; that economic media content does not Granger-cause individuals' opinions and expectations about their private economic situation. We accept the first hypothesis; that the economic media is not neutral and therefore publishes unequally many items with positive and negative sentiment. We find that the economic media reports slightly more items with positive than negative sentiment which differs to that found in the previous literature, which has shown negative reporting to be more common than positive reporting (Fogarty 2005; Arango-Kure *et al*. 2014; Boydstun *et al.* 2018). These different results may be explained by the fact that Swedish media is different to media in other economies and that we have more data stemming from a larger set of media outlets compared to other studies.

We provide evidence that individuals' opinions and expectations about their own personal economic situation are not Granger-caused by the economic media sentiment index. This implies that individuals use other sources of information for forming opinions about their own economic situation. Individuals may do this because they have access to other sources of information that are more relevant for their own personal economic situation. Alternatively, individuals may be overconfident in their opinions and expectations concerning their own economic situation in relation to the nationwide situation, which may cause individuals to ignore information about macroeconomic conditions that is published by the media.

Our results are in line with some of the previous literature, such as Hester and Gibson (2003), Boydstun *et al.* (2018) and Lamla *et al.* (2019), which have found that the public's perception of general economic conditions is affected by economic media, controlling for macroeconomic data. This indicates that the way the media frames news about economic conditions matters for individuals' perceptions about economic conditions. The general picture is thus that individuals' perception of macroeconomic conditions is a reflection of the information presented to them by the media. This is most likely due to individuals having few other information sources available to them that they can use to form an opinion about macroeconomic conditions. The economic media sentiment index that we have developed in this paper may thus be useful for forecasters who use data on individuals' opinions about macroeconomic conditions in their models for forecasting monthly macroeconomic activity.

# References

Almenberg, J., Dreber, A. (2015) 'Gender, stock market participation and financial literacy', *Economics Letters*, **137**, 140–142.

Altissimo, F., Cristadoro, R., Forni, M., Lippi, M., Veronese, G. (2010) 'New EUROCOIN: Tracking Economic Growth in Real Time', *The Review of Economics and Statistics*, **92**, 1024–1034.

Alwathainani, A.M. (2010) 'Does bad economic news play a greater role in shaping investors' expectations than good news?', *Global Economy and Finance Journal,* **3**, 27–43.

Andrew, B.C. (2007) 'Media-generated Shortcuts: Do Newspaper Headlines Present Another Roadblock for Low-Information Rationality?', *Press/Politics,* **12**, 24–43.

Arango-Kure, M., Garz, M., Rott, A. (2014) 'Bad News Sells: The Demand for News Magazines and the Tone of Their Covers', *Journal of Media Economics*, **27**, 199–214.

Arenius, P., Minniti, M., (2005) 'Perceptual Variables and Nascent Entrepreneurship', *Small Business Economics,* **24**, 233–247.

Armelius, H., Hull, I., Stenbacka Köhler, H. (2017), 'The Timing of Uncertainty Shocks in a Small Open Economy', *Economics Letters,* **155**, 31–34.

Baker, M., Wurgler, J. (2007) 'Investor Sentiment in the Stock Market', *Journal of Economic Perspectives*, **21**, 129–151.

Baker, S.R., Bloom, N., Davis, S.J. (2016) 'Measuring Economic Policy Uncertainty', *Quarterly Journal of Economics,* **131**, 1593–1636.

Bermingham, A., Smeaton A.F. (2010) 'Classifying sentiment in microblogs: Is brevity an advantage?', *Proceedings of the 19th ACM International Conference on Information and Knowledge Management. ACM, New York*, NY, USA, 1833–1836.

Birz, G., Lott, J.R. (2011) 'The effect of macroeconomic news on stock returns: New evidence from newspaper coverage', *Journal of Banking and Finance*, **35**, 2791–2800.

Blinder, A., Goodhart, C., Hildebrand, P., Lipton, D., Wyplosz, C. (2001) *How Do Central Banks Talk?,* Oxford; Information Press.

Bovi, M. (2009) 'Economic versus psychological forecasting. Evidence from consumer confidence surveys', *Journal of Economic Psychology,* **30**, 563–574.

Boydstun, A., Highton, B., Linn, S. (2018) 'Assessing the Relationship between Economic News Coverage and Mass Economic Attitudes', *Political Research Quarterly,* **71**, 989–1000.

Brown, S. & Taylor, K. (2006). 'Financial Expectations, Consumption and Saving: A Microeconomic Analysis', *Fiscal Studies*, **27**, 313–338.

Burnside, C., Han, B., Hirshleifer, D., Wang, Y.T. (2011) 'Investor Overconfidence and the Forward Premium Puzzle', *Review of Economic Studies*, **78**, 523–558.

Carroll, C.D (2003) 'Macroeconomic Expectations of Households and Professional Forecasters', *The Quarterly Journal of Economics*, **118**, 269–298.

Claessens, S., Dell'Ariccia, G., Igan, D., Laeven, L. (2010) 'Cross-country experiences and policy implications from the global financial crisis', *Economic Policy*, **4**, 267–293.

Doms, M.E., Morin, N.J. (2004) 'Consumer Sentiment, the Economy, and the News Media' *FRB of San Francisco*, Working Paper No. 2004-09.

Dräger, L. (2015), 'Inflation Perceptions and Expectations in Sweden – Are Media Reports the Missing Link?', *Oxford Bulletin of Economics and Statistics,* **77**, 681–700.

Ekeblom, D. (2014) 'Essays in Empirical Expectations', Lund University, PHD thesis.

Fama, E.F. (1990) 'Stock returns, expected returns, and real activity', *Journal of Finance,* **45**, 1089–1108.

Fogarty, B.J. (2005) 'Determining Economic News Coverage', *International Journal of Public Opinion Research,* **17**, 149–172.

Friedman, M. (1962) 'The Interpolation of Time Series by Related Series', *Journal of the American Statistical Association*, **57,** 729–757.

Gentzkow, M., Kelly, B., Taddy, M. (2019) 'Text as Data', *Journal of Economic Literature*, **57**, 535–574.

Gentzkow, M., Shapiro, J.M. (2008) 'Competition and Truth in the Market for News', *Journal of Economic Perspectives*, **22**, 133–154.

Goidel, K., Procopio, S., Terrell, D., Wu, H. D. (2010) 'Sources of Economic News and Economic Expectations' *American Politics Research*, **38**, 759–777.

Granger, C.W.J. (1969) 'Investigating Causal Relations by Econometric Models and Cross-spectral Methods', *Econometrica*, **37**, 424–438.

Gylfason, T. (1999) 'Exports, Inflation and Growth', *World Development*, **27**, 1031–1057.

Hester, J.B., Gibson, R. (2003) 'The Economy and Second-Level Agenda Setting: A Time-Series Analysis of Economic News and Public Opinion about the Economy', *Journalism and Mass Communications Quarterly,* **80**, 73–90.

Hirshleifer, D. (2001) 'Investor Psychology and Asset Pricing', *The Journal of Finance*, **56**, 1533–1597.

Hobolt, S.B., Leblond, P. (2009) 'Is My Crown Better than Your Euro?: Exchange Rates and Public Opinion on the European Single Currency', *European Union Politics*, **10**, 202–225.

Jaimovich, N., Rebelo, S. (2009) 'Can News about the Future Drive the Business Cycle?', *The American Economic Review*, **99**(4), 1097–1118.

Jain T.I., Nemade D. (2010) 'Recognizing Contextual Polarity in Phrase-Level Sentiment Analysis', *International Journal of Computer Applications* **7**, 0975–8887.

Johnson, M.A. Lamdin, D.J. (2012) 'Changes in Gasoline Prices and Consumer Sentiment', *Journal of Applied Business and Economics*, Forthcoming.

Keynes, J.M. (1936) *The General Theory of Employment, Interest and Money*. London, Macmillan.

Koellinger, P., Minniti, M., Schade, C. (2007) '"I think I can, I think I can'': Overconfidence and entrepreneurial behaviour', *Journal of Economic Psychology,* **28**, 502–527.

Lachmann, L.M., (1943) 'The Role of Expectations in Economics as a Social Science', *Economica* **10**, 12–23.

Lamla, M.J. & Lein, S.M. (2014) 'The Role of Media for Consumers' Inflation Expectation Formation', *Journal of Economic Behavior and Organization,* **106**, 62–77.

Lamla, M.J., Lein, S.M., Sturm, J.-E. (2019) 'Media Reporting and Business Cycles: Empirical Evidence based on News Data', *Empirical Economics,* forthcoming.

Lamla, M.J., Maag, T. (2012) 'The Role of Media for Inflation Forecast Disagreement of Households and Professional Forecasters', *Journal of Money, Credit and Banking*, **44**, 1325–1350.

Larcinese, V., Puglisi, R., Snyder J.M. (2011) 'Partisan bias in economic news: Evidence on the agenda-setting behavior of U.S. newspapers', *Journal of Public Economics*, **95,** 1178–1189.

Manning, C.D., Raghavan, P., Schütze H. (2010) *Introduction to Information Retrieval*, Cambridge University Press*,* Cambridge, UK.

Maziarz, M. (2015) 'A review of the Granger-causality fallacy', *The Journal of Philosophical Economics*, **8**, 86–105.

McCarthy, K.J., Dolfsma, W. (2014) 'Neutral Media? Evidence of Media Bias and its Economic Impact', *Review of Social Economy*, **72**, 42–54.

Mitchell, J., Smith, R.J., Weale, M.R., Wright, S., Salazar, E.L., (2005) 'An Indicator of Monthly GDP and an Early Estimate of Quarterly GDP Growth', *The Economic Journal*, **115**, 108–129.

Moretti, E. (2011) 'Social Learning and Peer Effects in Consumption: Evidence from Movie Sales', Review of Economic Studies, **78**, 356–393.

Mullainathan, S., Shleifer A. (2005) 'The Market for News', *American Economic Review*, **95**, 1031–1053.

Muth, J.F. (1961) 'Rational Expectations and the Theory of Price Movements', *Econometrica,* **29**, 315–335.

Nadeu, R., Niemi, R.G., Fan, D.P., Amato, T. (1999) 'Elite Economic Forecasts, Economic News, Mass Economic Judgement, and Presidential Approval', *The Journal of Politics*, **61**, 109–135.

Newey, W.K., West, K.D. (1987) 'A simple, positive semi-definite, heteroskedasticity and autocorrelation consistent covariance matrix', *Econometrica*, **55**: 703–708.

Ng, A.Y., Jordan, M.I. (2002) 'On Discriminative vs. Generative classifiers: A comparison of logistic regression and naïve Bayes', *Electronic Proceedings of the Neural Information Processing Systems*, conference paper.

Nguyen, V.H., Claus, E. (2013) 'Good news, bad news, consumer sentiment and consumption Behavior', *Journal of Economic Psychology,* **39**, 426-438.

O'Hare, N., Davy, M., Bermingham, A., Ferguson, P., Sheridan, P., Gurrin, C., Smeaton, A.F. (2009) 'Topic-Dependent Sentiment Analysis of Financial Blogs', *Proceedings of the 1st International CIKM Workshop on Topic-sentiment Analysis for Mass Opinion ACM, New York, NY, USA,* 9-16.

Olsson, E., Nord, L.W., Falkheimer, J. (2015) 'Media Coverage Crisis Exploitation Characteristics: A Case Comparison Study', *Journal of Public Relations Research*, **27**, 158–174.

Palmqvist, S., Strömberg, L. (2004) 'Households' inflation opinions – a tale of two surveys' *Sveriges Riksbank Economic Review* 4, 23–42.

Pigou, A. (1927) *Industrial Fluctuations*, MacMillan, London.

Ravn, M.O., Uhlig, H. (2002) 'On Adjusting the Hodrick-Prescott Filter for the Frequency of Observations', *The Review of Economics and Statistics*, **84**, 371–376.

Schwert, G.W. (1990) 'Stock Returns and Real Activity: A Century of Evidence', *The Journal of Finance*, **45**, 1237–1257.

Sebastiani, F. (2002) 'Machine Learning in Automated Text Categorization', *ACM Computing Surveys*, **34**, 1–47.

Serrano-Guerrero, J., Olivas, J.A., Romero, F.P., Herrera-Viedma, E. (2015) 'Sentiment analysis: A review and comparative analysis of web services', *Information Sciences*, 311, 18–38.

Sims, C.A. (2001) 'Implications of rational inattention', *Journal of Monetary Economics*, **50**, 665–690.

Soroka, N.S. (2006) 'Good News and Bad News: Asymmetric Responses to Economic Information', *The Journal of Politics*, **68**, 372–385.

Tannenbaum, P.H. (1953) 'The Effect of Headlines on the Interpretation of News Stories', *Journalism Bulletin*, **30**, 189-197.

van Raaij, W.F. (1989) 'Economic News, Expectations and Macroeconomic Behaviour', *Journal of Economic Psychology,* **10**, 473–493.

van Raaij, W.F., Gianotten, H.J. (1990) 'Consumer Confidence, Expenditure, Saving and Credit', *Journal of Economic Psychology,* **11**, 269–290.

Wåhlberg, A., Sjöberg, L. (2000) 'Risk perception and the media', *Journal of Risk Research*, **3**, 31–50.

Wilson, T., Wiebe, J., Hoffman, P. (2005) 'Recognizing Contextual Polarity in Phrase-Level Sentiment Analysis', *Proceedings of Human Language Technology Conference and Conference on Empirical Methods in Natural Language Processing,* 347–354.

Winseck, D. (2008) 'The State of Media Ownership and Media Markets: Competition or Concentration and Why Should We Care?', *Sociology Compass*, **2**, 34–47.

**Appendix**

**Table 9. Contribution of Information about Economic Variables to the Manual Sentiment Classification of Media Items**

| Variable ↑(↓) | Sentiment Contribution |
|---|---|
| Unemployment | Negative (Positive) |
| GDP | Positive (Negative) |
| SEK/EUR | Context Dependent (Context Dependent) |
| Stock prices | Positive (Negative) |
| Oil price | Negative (Positive) |
| Inflation | Context Dependent (Context Dependent) |

The symbols ↑(↓) represent information about an increase (decrease) in the variable. No distinction is made between increases in levels and, for example, percentage rates. The interpretation of the ↑(↓) symbols for the SEK/EUR variable is a depreciation (appreciation) of the SEK in relation to other currencies. In this table, the contributions are isolated, *i.e.* keeping everything else constant. In practice, however, most documents contain information about several of these variables. Sentiment classification is carried out by the author based on the full textual content of each media item.

**Table 10. Words with Largest difference in estimated conditional probability between positive and negative sentiment classes.**

| Negative Sentiment | | Positive Sentiment | |
|---|---|---|---|
| **English** | **Swedish** | **English** | **Swedish** |
| Financial | *Finansiella* | Percent | *Procent* |
| Government | *Regeringen* | Year | *År* |
| Decreased | *Sjönk* | millions of SEK | *MSEK* |
| American | *Amerikanska* | Amounted | *Uppgick* |
| Billions | *Miljarder* | Unemployment | *Arbetslösheten* |
| When | *När* | Good | *Bra* |
| Lower | *Sänker* | Continued | *Fortsatt* |
| Had | *Hade* | Strong | *Stark* |
| Points | *Punkter* | Repo rate | *Reporäntan* |
| Dollar | *Dollar* | Swedish | *Svensk* |
| Other | *Andra* | Growth | *Tillväxt* |
| Interventions | *Åtgärder* | The growth | *Tillväxten* |
| Weaker | *Svagare* | Increases | *Ökar* |
| Fiscal | *Finanspolitiska* | Inflation | *Inflationen* |
| Reduced | *Sänkt* | Expected | *Väntas* |
| Governments' | *Regeringens* | Will | *Kommer* |
| Write | *Skriver* | The year | *Året* |
| Recession | *Lågkonjunkturen* | Growing | *Växer* |
| Large | *Stora* | Forecast | *Prognos* |

An example of a media item with positive sentiment, positive words presented by table 10 are highlighted with grey background and negative words are highlighted with black background and white letters.

**Magdalena Andersson: Signaler om stark svensk ekonomi**
-----------------------------------------------------------------------------------------

Media: Affärsvärlden
Datum: 2017-04-10, 16:37:00
Publiceringsställe: webb

Vi har fått många positiva signaler om att svensk ekonomi står starkt och positiva utfall om de offentliga finanserna, från både SCB och ESV. Det sade finansminister Magdalena Andersson till Nyhetsbyrån Direkt vid Socialdemokraternas partikongress i Göteborg på måndagen.

"Det är oerhört positivt att den aktiva politik regeringen bedrivit för att få fler i jobb, sänka kostnaderna för migration och föra en stram finanspolitik, att det också gett resultat ", sade hon.

Hur står sig regeringens prognoser från i december?

"Det kommer jag att berätta på tisdag. Det är marknadspåverkande information och svarar jag nu då fethamnar jag i KU", sade hon.

I sin prognosuppdatering den 20 december var regeringens BNP-prognos 2,4 procent för 2017 och 1,8 procent för 2018, medan det offentliga sparandet väntades visa ett underskott på 0,4 procent av BNP i år och ett överskott på 0,2 procent nästa år.

ESV:s senaste prognos, från den 6 april, visade ett överskott i det finansiella sparandet på 0,3 procent i år och 0,5 procent nästa år.

**Search query used in Retriever database:**

"ekonomi" AND ("prognos" OR "rapport") AND ("affärsängel" OR "affärsängelnätverk" OR "aggregering" OR "aktie" OR "aktiebolag" OR "aktieindex" OR "aktiekurs" OR "aktivitetsindex" OR "allokering" OR "alternativkostnad" OR "amortering" OR "antidumpningstullar" OR "antidumpningsåtgärder" OR "anzcerta" OR "apec" OR "appreciering" OR "arbetsför ålder" OR "arbetsgivaravgift" OR "arbetsgivare" OR "arbetsgivarorganisation" OR "arbetskostnad" OR "arbetskraft" OR "arbetskraftsdeltagande" OR "arbetskraftsintensiv" OR "arbetskraftsinvandring" OR "arbetslös" OR "arbetslöshet" OR "arbetslöshetskassa" OR "arbetsmarknaden" OR "arbetsmarknadsavgift" OR "arbetsmarknadspolitiska åtgärder" OR "arbetsproduktivitet" OR "arbetstagare" OR "arbitrage" OR "asean" OR "attac" OR "avgift" OR "avkastning" OR "avräkning" OR "avskrivning" OR "avsättning" OR "avtalsförsäkringar" OR "avtalspension" OR "avtalsperiod" OR "avtalsrörelse" OR "arbetsför befolkning" OR "allmän jämvikt" OR "allmän löneavgift" OR "arbetsmarknadspolitiska program" OR "baisse" OR "balansräkning" OR "basarekonomi" OR "basbelopp" OR "basindustri" OR "basränta" OR "bebyggelsesektorn" OR "belöningssystem" OR "bergvärme" OR "beställningspunkt" OR "betalningsbalans" OR "bidrag" OR "bifirma" OR "bilaterala överenskommelser" OR "biobränsle" OR "biomassa" OR "blandekonomi" OR "blankning" OR "blockad" OR "bni" OR "bnp" OR "bojkott" OR "bolagsskatt" OR "bolagsstyrning" OR "boom" OR "bransch" OR "brp" OR "bruttoinkomst" OR "bruttoinvestering" OR "bruttolön" OR "bruttonationalprodukt" OR "bruttoskuld" OR "brytpunkt" OR "bubbla" OR "budget" OR "budgeteringsmarginal" OR "budgetproposition" OR "budgetsaldo" OR "budgetunderskott" OR "bytesbalans" OR "bytesförhållande" OR "börs" OR "beskattningsbar förvärvsinkomst" OR "bnp-gap" OR "budgetpolitiska mål" OR "bnp nominell" OR "bnp potentiell" OR "bnp real" OR "bank run" OR "cdm" OR "centralbank" OR "chock" OR "clrtap" OR "cofog" OR "cotonou-avtalet" OR "csr" OR "cykel" OR "cykliska bolag" OR "corporate governance" OR "cdm/ji" OR "ceteris paribus" OR "corporate social responsibility" OR "dagslåneräntan" OR "decilen" OR "deflatera" OR "deflation" OR "depreciering" OR "depression" OR "derivat" OR "detaljhandel" OR "devalvering" OR "direktinvesteringar" OR "diskontera" OR "doharundan" OR "dotterbolag" OR "dumping" OR "duopol" OR "direktverkande el" OR "dirty floating" OR "disponibel inkomst" OR "dow jones index" OR "direkta skatter" OR "dualt skattesystem" OR "ebita" OR "ecb" OR "ees" OR "effekt" OR "effektiva växelkurser" OR "effektivitet" OR "efta" OR "efterfrågan" OR "efterfrågeinflation" OR "efterfrågepolitik" OR "egenavgifter" OR "ekonometri" OR "ekonomiskt bistånd" OR "elasticitet" OR "elcertifikat" OR "elektricitet" OR "elproduktion" OR "elsimulator" OR "embargo" OR "emission" OR "empiri" OR "emu" OR "energi" OR "energi" OR "energibalans" OR "energimått" OR "enmansföretag" OR "entreprenör" OR "entropi" OR "erm2" OR "ersättningsgrad" OR "ersättningsinvestering" OR "etableringsuppdraget" OR "eu" OR "eu-15" OR "eu-19" OR "eu-25" OR "eu-27" OR "euro" OR "europeiska centralbanken" OR "eurostat" OR "export" OR "externa effekter" OR "enskild firma" OR "effektiv ränta" OR "enkel ränta" OR "ekonomisk standard" OR "europeiska unionen" OR "fackförbund" OR "fast växelkurs" OR "fastighetsskatt" OR "fattigdomsstrecket" OR "federalism" OR "finanser" OR "finansiell balans" OR "finansmarknad" OR "finanspolitik" OR "flexicurity" OR "fluktuera" OR "flytande valuta" OR "fond" OR "forward" OR "fossila bränslen" OR "fou" OR "fredsplikt" OR "frihandel" OR "frihandelsområde" OR "friktionsarbetslöshet" OR "friskola" OR "frivilliga exportbegränsningar" OR "fåmansbolag" OR "förbundsavtal" OR "företag" OR "förmögenhetsskatt" OR "förnyelsebar energi" OR "försörjningsbalans" OR "försörjningsbörda" OR "förvärvsarbetande" OR "förvärvsfrekvens" OR "förädlingsindustri" OR "förädlingsvärde" OR "fast kostnad" OR "flexibla mekanismer" OR "fysiska personer" OR "fasta priser" OR "fri prissättning" OR "fiskala skatter" OR "finansiellt sparande" OR "full sysselsättning" OR "g7(8)" OR "gats" OR "gatt" OR "gdp" OR "generalindex" OR "generella bidrag" OR "geotermisk energi" OR "giga" OR "globalisering" OR "glokalisering" OR "grossist" OR "grundavdrag" OR "guldmyntfot" OR "göteborgsprotokollet" OR "greenfield investering" OR "giips-länderna" OR "grön skatteväxling" OR "handelsbalans" OR "handelshinder" OR "handelsintensitet" OR "handelskammare" OR "handelsnetto" OR "handelspolitik" OR "hausse" OR "hedgefond" OR "humankapital" OR "husarbete" OR "hysteresis" OR "hängavtal" OR "högkonjunktur" OR "hdi human development index" OR "handels- och kommanditbolag" OR "handelsbalansens saldo" OR "icc" OR "ilo" OR "imf" OR "immaterialrätt" OR "import" OR "importrestriktion" OR "index" OR "industri" OR "inferiora varor"

OR "inflation" OR "inflation hyper" OR "inflationsmål" OR "infrastruktur" OR "inkomst" OR "inkomstelasticitet" OR "inkurans" OR "inlåning" OR "innovation" OR "innovatör" OR "insourcing" OR "intraprenör" OR "intäkt" OR "investering" OR "investeringskvot" OR "investmentbolag" OR "itc" OR "i-land" OR "indirekta skatter" OR "immateriell tillgång" OR "inflation underliggande" OR "ji" OR "jobbavdraget" OR "jobbskatteavdraget" OR "jobless growth" OR "joule" OR "jämvikt" OR "juridiska personer" OR "jobba svart" OR "kalenderkorrigerad" OR "kapacitetsutnyttjande" OR "kapital" OR "kapitalbalans" OR "kapitalbeskattning" OR "kapitalflöde" OR "kapitalförslitning" OR "kapitalinkomstskatt" OR "kapitalism" OR "kapitalist" OR "kapitalskatt" OR "kapitalstock" OR "kartell" OR "kennedyrundan" OR "keynesianism" OR "kilo" OR "kiloton" OR "kilowattimme" OR "klassisk arbetslöshet" OR "koks" OR "koldioxidekvivalenter" OR "koldioxidintensitet" OR "kollektiva varor" OR "kollektivavtal" OR "kommandoekonomi" OR "kommunal inkomstskatt" OR "kommunalförbund" OR "kommunförbund" OR "kommunism" OR "komparativa fördelar" OR "kompetens" OR "koncern" OR "kondenskraftverk" OR "konjunktur" OR "konjunkturcykler" OR "konkurrens" OR "konkurrenskraft" OR "konkurs" OR "konsolidera" OR "konsumtionsutgifter" OR "kontraktiv" OR "konvergenskrav" OR "kooperativ" OR "korspriselasticitet" OR "kostnad" OR "kpi" OR "kraftvärmeverk" OR "kredit" OR "kreditförlust" OR "kreditinstitut" OR "kreditmarknad" OR "kvot" OR "kväveoxider" OR "kyoto-överenskommelsen/kyotoprotokollet" OR "kärnkraft" OR "köp av verksamhet" OR "köpkraft" OR "köpkraftsparitet" OR "kollektiva nyttigheter" OR "kapitalintensiv produktion" OR "kort ränta" OR "konkurrensutsatt sektor" OR "kommunala sektorn" OR "kort sikt" OR "lagerförändringar" OR "lagerinvesteringar" OR "las" OR "ledande indikatorer" OR "liberalisering" OR "likviditet" OR "likviditetsfällan" OR "likviditetsöverskott" OR "lissabonstrategin" OR "lockout" OR "lou" OR "lov" OR "lägstalön" OR "lågkonjunktur" OR "låneinstitut" OR "lön" OR "lönebildning" OR "lönebildning lokal" OR "löneglidning" OR "lönepolicy" OR "lönerevision" OR "lönesamtal" OR "löneskatt" OR "lönespiral" OR "lönespridning" OR "lönestrukturen" OR "löneutveckling" OR "löntagarfonder" OR "löptid" OR "likvida medel" OR "löpande priser" OR "lång ränta" OR "lång sikt" OR "lex uggla" OR "maastrichtkriterierna" OR "maastrichtskulden" OR "mai" OR "makroekonomi" OR "marginalavkastning" OR "marginalinkomst" OR "marginalintäkt" OR "marginalprodukt" OR "marginalskatt" OR "marginalskatten" OR "marknad" OR "marknadsandel" OR "marknadsekonomi" OR "marknadsmisslyckande" OR "marknadsränta" OR "marknadsvärde" OR "medarbetarsamtal" OR "medelstora företag" OR "median" OR "medianlön" OR "medlingsinstitutet" OR "mega" OR "megawattimme" OR "merchanting" OR "mercosur" OR "mgn" OR "mikroekonomi" OR "mikroföretag" OR "minimilön" OR "marketing mix" OR "moms" OR "monetär bas" OR "monopol" OR "monopsoni" OR "mul" OR "multilateral" OR "multinationell" OR "märket" OR "materiell tillgång" OR "nafta" OR "nationalekonomi" OR "nationalräkenskaper" OR "naturgas" OR "nettoinkomst" OR "nettoinvestering" OR "nettolön" OR "nettomarginal" OR "ngos" OR "nominell" OR "nominell lön" OR "nominellt värde" OR "nr" OR "nyemission" OR "näringsfrihet" OR "näringslivet" OR "näringsverksamhet" OR "nådiga luntan" OR "nominell ränta" OR "obligation" OR "oecd" OR "offentlig konsumtion" OR "offshoring" OR "ohälsotalet" OR "oktroj" OR "oligopol" OR "omsättning" OR "opec" OR "outsourcing" OR "ozon" OR "offentlig sektor" OR "offentlig upphandling" OR "parter" OR "partihandel" OR "patent" OR "pengar" OR "penningmängd" OR "penningpolitiken" OR "pension avgiftsbestämd" OR "pension förmånsbestämd" OR "pensionsgrundande inkomster" OR "per capita" OR "percentil" OR "periodisering" OR "peta" OR "piigs-länderna" OR "planekonomi" OR "plusjobb" OR "potentiell bnp" OR "ppp" OR "premieobligation" OR "prestationslön" OR "primärenergi" OR "primärvård" OR "prisbasbelopp" OR "priselasticitet" OR "prismekanism" OR "prisstabilitet" OR "privat konsumtion" OR "privatisering" OR "produktionsfaktorer" OR "produktivitet" OR "prognos" OR "protektionism" OR "punktskatt" OR "platt skatt" OR "progressiv skatt" OR "proportionell skatt" OR "p/e-tal" OR "p/s-tal" OR "quad" OR "rationalisering" OR "real" OR "real växelkurs" OR "realekonomisk fördelning" OR "realkapital" OR "reallön" OR "realränta" OR "reavinst" OR "recession" OR "referensränta" OR "referensår" OR "regleringsbrev" OR "relativ kostnad" OR "remburs" OR "repa" OR "reporäntan" OR "reservationslön" OR "restavgift" OR "resultat" OR "resultatlön" OR "resursutnyttjande" OR "revalvering" OR "revers" OR "riksbanken" OR "riksgälden" OR "riskkapitalbolag" OR "rot-avdrag" OR "rullande budget" OR "rut-avdrag" OR "ränta" OR "ränteavdrag" OR "räntepunkt" OR "ränterisk" OR "råvara" OR "rörlig lön" OR "rörlig växelkurs" OR "saldo" OR "saltsjöbadsavtalet" OR "samhällsekonomi" OR "scb" OR "skatt på arbete" OR "skatt på kapital" OR "skattebas" OR "skattekil" OR "skattekvot" OR "skattetryck" OR "skift" OR "skiktgräns" OR "skolpeng" OR "skuldkvot" OR "sme-företag" OR "smed" OR "små företag" OR "social dumping" OR "social omsorg" OR "sociala avgifter" OR "sociala klausuler" OR "socialbidrag" OR "socialisering" OR "socialism" OR "solidarisk lönepolitik" OR "soliditet" OR "somatisk vård" OR "sparkvot" OR "spekulation" OR "stabiliseringspolitik" OR "stagflation" OR "stagnation" OR "statens lånebehov" OR "statens skatteinkomster" OR "statens skatteintäkter" OR "statsbudget" OR "statsbudgetens saldo" OR "statslåneräntan" OR "statsobligation" OR "statspapper" OR "statsskuld" OR "statsskuldsväxel" OR "stibor" OR "stordriftsfördelar" OR "strejk" OR "strejk vild" OR "strejkvarsel" OR "stridsåtgärd" OR "strukturell arbetslöshet" OR "strukturomvandling" OR "styrränta" OR "stämpelskatt" OR "subprimelån" OR "substitutionselasticitet" OR "subvention" OR "svart marknad" OR "svavel" OR "sympatiåtgärd" OR "sysselsättning" OR "sysselsättningsandel" OR "sysselsättningskvot" OR "sänkor" OR "säsongsrensad" OR "skatt på varor och tjänster" OR "statsbudgetens totala inkomster" OR "statsbudgetens totala utgifter" OR "taxerad förvärvsinkomst" OR "taxeringsvärde" OR "tcw-index" OR "teknikindustri" OR "tera" OR "terminskontrakt" OR "terrawattimme" OR "tidsserie" OR "tillväxt" OR "tjänstebalansens saldo" OR "tjänstehandelsintensitet" OR "tjänstepension" OR "tjänster" OR "tobinskatt" OR "tokyorundan" OR "totala skatteintäkter" OR "terms of trade" OR "transaktionskostnader" OR "transferering" OR "transnationella företag" OR "trend" OR "trips" OR "trängselskatt" OR "ttip" OR "tull" OR "tullunion" OR "tvistelösning" OR "utgiftstak" OR "utlandsskuld" OR "utträngning" OR "utvecklingssamtal" OR "u-land" OR "unctad" OR "underliggande budgetsaldo" OR "unilaterala handlingar" OR "upphandling" OR "uruguayrundan" OR "utanförskap" OR "utbildning" OR "utbud" OR "utbudspolitik" OR "utgift" OR "verkningsgrad" OR "valuta" OR "valutakurs" OR "valutareserv" OR "valutaskuld" OR "valutaunion" OR "valutautflöde" OR "varor" OR "varsel" OR "verifikation" OR "vinst" OR "voc" OR "vårdval" OR "välfärd" OR "välstånd" OR "världsbanken" OR "världsekonomi" OR "världshandel" OR "värnskatt" OR "värnskatten" OR "växelkurs" OR "växthuseffekten" OR "växthusgaser" OR "världs-bnp" OR "wto" OR "ålderspensionssystemet" OR "årslön" OR "årsredovisning" OR "återförsäljare" OR "övergödning" OR "överhettning" OR "överkapacitet" OR "öppet arbetslösa" OR "öppna marknadsoperationer").

**Table 11. Examples of 10 Media Items Classified as Positive Sentiment by the Author**

| Date | Source | Published | Headline in Swedish | Headline in English |
|---|---|---|---|---|
| 2014-02-05 10:49 | Veckans Affärer | Online | *Sveriges ekonomi tar fart igen* | *Sweden's economy is gaining momentum again* |
| 1999-04-28 | Helsingborgs Dagblad | Print | *Nordbanken spår ny köpfest Prognos flaggar för fyra feta år fram till 2002* | *Nordbanken predicts new spending spree: Forecasts four good years until 2002* |
| 2015-02-25 10:30 | Affärsvärlden | Online | *Bygg: BI höjer prognos för bygginvesteringar Sverige 2014-2016* | *Building: BI raises forecasts for construction investments Sweden 2014-2016* |
| 2010-03-01 08:18 | Sydsvenskan | Online | *SEBA: SEB:s Företagarpanel om konjunkturen: En majoritet av företagarna...* | *SEBA: SEB's Business Panel on the business cycle: A majority of entrepreneurs ...* |
| 2004-08-31 13:11 | Helsingborgs Dagblad | Online | *SEB spår goda tider* | *SEB predicts good times* |
| 2004-12-20 10:06 | Aktiespararna | Online | *KI: investeringarna bidrar till hög tillväxt...* | *KI: Investments contribute to strong growth...* |
| 1995-08-31 13:26 | TT Nyhetsbyrån | Print | *Nordbanken tror på Sveriges Ekonomi* | *Nordbanken believes in the Swedish economy* |
| 2006-12-20 10:56 | Sveriges radio ekot | Online | *KI tror på svensk ekonomi* | *KI believes in the Swedish economy* |
| 2013-04-22 21:51 | Metro | Online | *Nya ljusglimtar för svensk ekonomi* | *New silver linings for the Swedish economy* |
| 2000-02-26 | Dagens Nyheter | Print | *Optimistisk prognos av Handelsbanken* | *Optimistic forecast by Handelsbanken* |

**Table 11. Examples of 10 media items classified as negative sentiment by the author.**

| Date | Source | Published | Headline in Swedish | Headline in English |
|---|---|---|---|---|
| 1999-04-30 | Svenska Dagbladet | Print | *Ekonomin tappar fart* | *The economy is losing momentum* |
| 2009-02-25 10:35 | Privata Affärer | Online | *Byggindustrierna: Investeringar -5% i år* | *Construction industries: Investments -5% this year* |
| 2006-09-22 10:37 | allehanda.se | Online | *Börsen ned på bred front* | *Stock market down broadly* |
| 2011-11-10 09:48 | Hallands Nyheter | Online | *Rehn: `Risk för ny recession`* | *Rehn: 'Risk for new recession'* |
| 2008-05-30 10:39 | Ystads Allehanda | Online | *Tillväxten lägre än väntat* | *Growth lower than expected* |
| 2011-09-06 10:25 | Borås Tidning | Online | *ESV: Risk för tvärnit i ekonomin* | *ESV: Risk of sudden halt in the economy* |
| 2011-08-11 10:40 | Dagens Industri | Online | *Borg skriver ned BNP-prognosen påtagligt* | *Borg writes down the GDP forecast significantly* |
| 2014-07-23 18:09 | Kristianstadsbladet | Online | *IMF sänker sin USA-prognos – igen* | *The IMF is lowering its US forecast – again* |
| 2011-03-28 09:49 | Svenska Dagbladet | Online | *Krisen långt från över i Europa* | *The crisis far from over in Europe* |
| 2009-02-26 | nyheter24 | Online | *Sveriges BNP väntas störtdyka* | *Sweden's GDP is expected to plummet* |

---

[1] We use a search query which utilizes the Boolean expressions AND, OR and NOT. An example of a simplistic Boolean search query that searches the Retriever database for media items that talk about the economy could be: *"GDP" AND "unemployment"*. This query will result in media items that contain at least one occurrence of both of the words "GDP" and "unemployment". See the Appendix for the actual query used and Losee (1997) for a detailed description of Boolean search queries.

[2] We filter our search on dates and language (Swedish) and require that the search engine searches through the full textual content of the media items.

[3] The test data consist of 500 media items corresponding to 0.28 % of the total amount of media items in the full data set.

[4] See the Appendix for lists with some examples of economic media items that were manually classified by the author.

[5] See Gentzkow *et al.* (2019) and Manning *et al.* (2010) for a more detailed description of NB. We use pseudo code which describes an algorithm for implementing NB in practice from Manning *et al.* (2010). We implement this algorithm in a C++ program which is available upon request.

[6] See Appendix for a list of our NB models' top 20 words with largest difference in probability between the positive- and negative sentiment classes and an example of a classified media item.

[7] We split the sample into two groups using the sources' average number of words per item and number of items in the data set as feature vectors. We weight items based on their distance (inverse L2-norm) to a set of examples when calculating a scaled EMSI-measure for the two groups respectively.

[8] NIER is a government agency that performs analyses and forecasts of the Swedish and international economy as a basis for economic policy in Sweden.

[9] For more details on the survey see Palmqvist & Strömberg (2004) and Ekeblom (2014).

[10] The ADF-test rejects the null at the 1-% level for sweNow, ownNow and ownFuture and at the 5% level for sweFuture.

[11] We use a Quandt likelihood ratio test for structural break in sweNow and find that the variable has a break at 1997. We thus manually level-correct the series at 1997.